\documentclass[aps,onecolumn,superscriptaddress]{revtex4}
\usepackage{times}
\usepackage{color}
\usepackage{float}
\usepackage{amssymb}
\usepackage{epsfig}
\usepackage{booktabs}
\usepackage{BOONDOX-cal}
\usepackage{mathrsfs}
\usepackage{amsmath}

\begin{document}

\title{Quantum corrections to the entropy in a driven quantum Brownian motion model}

\author{Tian Qiu}
\affiliation{School of Physics, Peking University, Beijing, 100871, China}

\author{H. T. Quan}\thanks{Corresponding author: htquan@pku.edu.cn}
\affiliation{School of Physics, Peking University, Beijing, 100871, China}
\affiliation{Collaborative Innovation Center of Quantum Matter, Beijing 100871, China}
\affiliation{Frontiers Science Center for Nano-optoelectronics, Peking University, Beijing, 100871, China}

\date{\today}

\begin{abstract}
Quantum Brownian motion model is a typical model in the study of nonequilibrium quantum thermodynamics. Entropy is one of the most fundamental physical concepts in thermodynamics. In this work, by solving the quantum Langevin equation, we study the von Neumann entropy of a particle undergoing quantum Brownian motion. In both the strong and the weak coupling regimes, we obtain the analytical expression of the time evolution of the Wigner function in terms of the initial Wigner function. The result is applied to the thermodynamic equilibrium initial state, which reproduces its classical counterpart in the high-temperature limit. Based on these results, for those initial states having well-defined classical counterparts, we obtain the explicit expression of the quantum corrections to the entropy of the system. Moreover, under the Markovian approximation, we obtain the expression of the quantum corrections to the total entropy production rate ${e_{\rm p}}$ and the heat dissipation rate ${h_{\rm d}}$. Our results bring important insights to the understanding of entropy in open quantum systems.
\end{abstract}

\maketitle
\section{Introduction}
Quantum thermodynamics \cite{Wigner1932, Esposito2009, Campisi2011, Binder2018, Deffner2019} is an emerging field studying the nonequilibrium statistical mechanics of the quantum dissipative systems \cite{Yu1994, Breuer2002, Hanggi2005, Caldeira2014}. Quantum work \cite{Kurchan, Tasaki, Talkner2007, Talkner2016, Ken2018}, quantum heat \cite{Saito2007, Aurell2017, Ken2018b, Hu2018, Hu2020a, Hu2020b}, and quantum entropy production \cite{Aurell2015, Weiderpass2020, Pucci2013, Esposito2010} are among the most basic concepts, which play important roles in the study of work extraction and heat transfer in quantum devices, such as the quantum heat engines and refrigerators \cite{Scovil1959, Geusic1959, Alicki1979, Kosloff1984, Geva1992, Geva1996, Bender2000, Scully2003, Kieu2006, Quan2007, Allahverdyan2010, Linden2010, Lutz2014, Dong2015, Beau2016, Karimi2016}.

A typical exactly solvable model used for addressing these problems is the quantum Brownian motion model proposed by Caldeira and Leggett \cite{Caldeira1983, Caldeira2014}. It consists of a system described by the Hamiltonian \begin{small}${\hat{H}_S}$\end{small} (often a harmonic oscillator \cite{Hu1992}), a heat bath of harmonic oscillators with the Hamiltonian \begin{small}${\hat{H}_B}$\end{small}, and the interaction Hamiltonian \begin{small}${\hat{H}_{SB}}$\end{small}. One can analytically integrate out the degrees of freedom of the heat bath, which brings important insights to the understanding of the thermodynamics of open quantum systems. In the studies about entropy production and heat transfer, previous efforts have been focused mainly on the entropy production in the heat bath. For example, in Ref. \cite{Aurell2015}, by adapting the Feynman-Vernon influence functional formalism, the change of the von Neumann entropy of the heat bath is computed. In Refs. \cite{Oono1998, Sasa2001, Qian2001, Qian2001b, Seifert2005, Imparato2006, Ge2006, Ge2009, Saha2009, Ge2010, Esposito2010a, Esposito2010b, Esposito2010c, R2012, Ge2018}, the expressions of some basic nonequilibrium thermodynamic quantities, such as the total entropy production rate \begin{small}${e_{\rm p}}$\end{small}, the heat dissipation rate \begin{small}${h_{\rm d}}$\end{small}, the housekeeping heat \begin{small}${Q_{\rm hk}}$\end{small} and the excess heat \begin{small}${Q_{\rm ex}}$\end{small}, are discussed for a classical isothermal process described by a master equation or a Fokker-Planck (Langevin) equation. More recently, the energy exchange, and thus the entropy exchange, between the system and the heat bath is calculated in Refs. \cite{Ken2018b, Hu2020a, Hu2020b}. However, the quantum-classical correspondence, and especially the quantum corrections to the entropy production in the quantum Brownian motion model, as well as the total entropy production, have not been explored so far (but see Ref. \cite{Weiderpass2020}).

In this article, we study the time evolution of the quantum entropy and its corrections to the classical entropy in an open quantum system. Specifically, we calculate the von Neumann entropy in the quantum Brownian motion model subject to a driving force, as well as the total entropy production. The von Neumann entropy of a quantum system described by the density matrix \begin{small}${\hat{\rho}}$\end{small} is given by \cite{von Neumann1927, von Neumann1955}
\begin{small}
\begin{equation}\label{eq:0000}
S_q=-\rm{Tr}\left[\hat{\rho}\ln\hat{\rho}\right].
\end{equation}
\end{small}Here, we have set the Boltzmann's constant to be ${1}$, and thus the entropy becomes dimensionless. It is difficult to calculate the von Neumann entropy through its definition Eq. (\ref{eq:0000}), because one has to diagonalize an infinite-dimensional matrix in order to compute the trace of a function. However, by reformulating the problem in the phase space the calculation can be significantly simplified \cite{Qiu2020}. By solving the quantum Langevin equation exactly, we obtain the analytical expression of the Wigner function at an arbitrary time \begin{small}${t}$\end{small} in terms of the initial Wigner function, which is valid in both the strong and the weak coupling regimes. Then by adapting the method in Ref. \cite{Qiu2020}, for initial states which have well-defined classical counterparts, we find that if we expand the von Neumann entropy in powers of \begin{small}${\hbar}$\end{small}, the zeroth-order term reproduces the classical Gibbs entropy. We can also obtain the quantum correction to the entropy of the system and its time evolution. Moreover, under the Markovian approximation, we obtain the expression of the quantum corrections to the total entropy production rate ${e_{\rm p}}$ and the heat dissipation rate ${h_{\rm d}}$. Our results bring important insights to the understanding of the entropy in an open quantum system.

This article is organized as follows. We begin in Sec.\uppercase\expandafter{\romannumeral2} with a description of the model and the quantum Langevin equation. In Sec. III we derive the general solution to the quantum Langevin equation of the driven quantum Brownian motion model. In Sec. IV we obtain the expression of the reduced Wigner function of a driven quantum Brownian motion model. In Sec. V, by using the method developed in Ref. \cite{Qiu2020}, we calculate the quantum corrections to the entropy of the system and its dynamical evolution. In Sec. VI, based one the results in Sec. V, we obtain the expression of the quantum corrections to the total entropy production rate ${e_{\rm p}}$ and the heat dissipation rate ${h_{\rm d}}$. Finally, in Sec. VII we make some remarks and summarize our results.

\section{The model and the derivation of the quantum Langevin equation}
We consider the quantum Brownian motion described by the Caldeira-Leggett model \cite{Caldeira2014, Caldeira1983, Hu1992}. The system that we consider is a harmonic oscillator subject to a time-dependent driving force \begin{small}${\hat{f}(t)}$\end{small} \cite{Xu2009}. The system is linearly coupled to a heat bath consisting of a set of harmonic oscillators. The Hamitonian of the composite system is given by \begin{small}${\hat{H}_{\rm{tot}}(t)=\hat{H}_S(t)+\hat{H}_B+\hat{H}_{SB}}$\end{small}, with
\begin{small}
\begin{subequations}
\begin{eqnarray}\label{eq:0001} &&\hat{H}_S(t)=\frac{\hat{\mathcal{p}}^2(t)}{2 m_0}+\frac{1}{2}m_0\omega_0^2\left(\hat{\mathcal{q}}(t)-\frac{\hat{f}(t)}{m_0\omega_0^2}\right)^2,\ \ \ \ \ \ \ \\
&&\hat{H}_B=\sum^{N}_{j=1}\left(\frac{\hat{p}^2_j(t)}{2 m_j}+\frac{1}{2}m_j\omega_j^2\hat{q}_j^2(t)\right),\\
&&\hat{H}_{SB}=-\hat{\mathcal{q}}(t)\sum_{j=1}^{N}C_j\hat{q}_j(t)+\sum_{j=1}^{N}\frac{C_j^2}{2m_j\omega_j^2}\hat{\mathcal{q}}^2(t),
\end{eqnarray}
\end{subequations}
\end{small}where \begin{small}${m_0}$\end{small}, \begin{small}${\omega_0}$\end{small}, \begin{small}${\hat{\mathcal{q}}}$\end{small}, \begin{small}${\hat{\mathcal{p}}}$\end{small} and \begin{small}${m_j}$\end{small}, \begin{small}${\omega_j}$\end{small}, \begin{small}${\hat{q}_j}$\end{small}, \begin{small}${\hat{p}_j}$\end{small} are the mass, angular frequencies, coordinates and momenta of the system and the \begin{small}${j}$\end{small}th harmonic oscillator of the heat bath, respectively, and \begin{small}${C_j (j=1,2,3,...)}$\end{small} are the coupling constants. Here, we have included the counterterm \begin{small}${\sum_j (C_j^2/2m_j\omega_j^2)\hat{\mathcal{q}}^2(t)}$\end{small} in the interaction Hamiltonian to cancel the negative frequency shift of the potential \cite{Caldeira1983a}.

The equation of motion of the time-dependent operator can be obtained by using the Heisenberg equation
\begin{small}
\begin{equation}\label{eq:0002}
i\hbar\dot{\hat{O}}=[\hat{O},\hat{H}],
\end{equation}
\end{small}which gives the time derivative (denoted by the superposed dot) of an
arbitrary operator \begin{small}${\hat{O}}$\end{small}. Then we have \cite{Ford1988}
\begin{small}
\begin{equation}\label{eq:0003}
\left\{
\begin{aligned}
&\dot{\hat{\mathcal{q}}}\!=\!\frac{1}{m_0} \hat{\mathcal{p}} \\
&\dot{\hat{\mathcal{p}}}\!=\!-m_0\omega_0^2\hat{\mathcal{q}}\!+\!\hat{f}(t)\!+\!\sum^N_{j=1}\!C_n\!\left(\hat{q}_j\!-\!\frac{C_j}{m_j\omega_j^2}\hat{\mathcal{q}}\right)
\end{aligned}\ \
\right.
\end{equation}
\end{small}for the system, and\begin{small}
\begin{equation}\label{eq:0004}
\left\{
\begin{aligned}
&\dot{\hat{q}}_j=\frac{1}{m_j} \hat{p}_j \\
&\dot{\hat{p}}_j=-m_j\omega_j^2\left(\hat{q}_j-\frac{C_j}{m_j\omega_j^2}\hat{\mathcal{q}}\right)
\end{aligned}\ \ \ \ \ \ \ \ \ \ \ \ \ \ \ \ \ \ \ \ \ \ \ \ \ \ \
\right.
\end{equation}
\end{small}for the \begin{small}${j}$\end{small}th harmonic oscillator of the heat bath. Solving Eq. (\ref{eq:0004}) and substituting it into Eq. (\ref{eq:0003}), one can obtain the quantum Langevin equation of the driven Brownian particle \cite{Ford1988}
\begin{small}
\begin{equation}\label{eq:0005}
m_0 \ddot{\hat{\mathcal{q}}}(t)+\int^{t}_{0}dt' \mu(t-t')\dot{\hat{\mathcal{q}}}(t')+m_0\omega_0^2 \hat{\mathcal{q}}(t)+\mu(t)\hat{\mathcal{q}}(0)=\hat{F}(t)+\hat{f}(t),
\end{equation}
\end{small}where \begin{small}${\mu(t)}$\end{small} is the memory function and it is given by
\begin{small}
\begin{equation}\label{eq:0006}
\mu(t)=\sum_{j}\frac{C_j^2}{m_j\omega_j^2}\cos(\omega_j t),
\end{equation}
\end{small}and \begin{small}${\hat{F}(t)}$\end{small} is the fluctuating force operator and it can be expressed in terms of the initial bath variables
\begin{small}
\begin{equation}\label{eq:0007}
\hat{F}(t)=\sum_j C_j\left[\hat{q}_j(0)\cos(\omega_j t)+\hat{p}_j(0)\frac{\sin(\omega_j t)}{m_j \omega_j}\right].
\end{equation}
\end{small}It is straightforward to show that the correlation and the commutator can be expressed as \cite{Ford1988, Ford2001}
\begin{small}
\begin{equation}\label{eq:0008}
\frac{1}{2}\langle \hat{F}(t)\hat{F}(t')+\hat{F}(t') \hat{F}(t)\rangle=\frac{1}{\pi}\!\int^{\infty}_{0}\!d\omega\! \ \rm{Re}\{\tilde{\mu}(\omega\!+\!i0^+)\}\hbar\omega\coth{\!\left(\!\frac{\beta\hbar\omega}{2}\!\right)\!}\cos[\omega(t\!-\!t')],
\end{equation}
\begin{equation}\label{eq:0009}
\left[\hat{F}(t), \hat{F}(t')\right]=\frac{2\hbar}{i\pi}\!\int^{\infty}_{0}\!d\omega \ \rm{Re}\{\tilde{\mu}(\omega+i0^+)\}\omega\sin[\omega(t-t')],
\end{equation}
\end{small}where the bracket \begin{small}${\langle...\rangle}$\end{small} depicts the quantum expectation value, and \begin{small}${\tilde{\mu}(z)}$\end{small} is the Fourier transform of the memory function:
\begin{small}
\begin{equation}\label{eq:0010}
\tilde{\mu}(z)=\int^{\infty}_{0}dt\ \mu(t) e^{i z t}, \ \ \ \ \rm{Im}\ z>0.
\end{equation}\end{small}In the following, we will try to solve Eq. (\ref{eq:0005}) by using the Green function approach.

\section{General solution to the quantum Langevin equation (\ref{eq:0005}) }
For the stationary process, the system is held fixed at the origin in the distant past \cite{Ford1988}. From Eq. (\ref{eq:0005}), we obtain the quantum Langevin equation for the stationary process
\begin{small}
\begin{equation}\label{eq:0011}
m_0\ddot{\hat{\mathcal{q}}}^{(\!s\!)\!}(t)\!+\!\int^{t}_{-\infty}dt'\mu(t-t')\dot{\hat{\mathcal{q}}}^{(\!s\!)\!}(t')\!+\!m_0\omega_0^2\hat{\mathcal{q}}^{(\!s\!)\!}(t)\!=\!\hat{F}(t)\!+\!\hat{f}(t),
\end{equation}
\end{small}where \begin{small}${\hat{f}(t)}$\end{small} is switched on at \begin{small}${t=0}$\end{small}, and the solution can be written as \cite{Ford2001}
\begin{small}
\begin{equation}\label{eq:0012}
\hat{\mathcal{q}}^{(s)}(t)=\hat{\mathcal{q}}^{(F)}(t)+\hat{\mathcal{q}}^{(f)}(t),
\end{equation}
\end{small}where
\begin{small}
\begin{subequations}
\begin{eqnarray}\label{eq:0012a}
&&\hat{\mathcal{q}}^{(F)}(t)=\int^{t}_{-\infty}dt'\ G(t-t')\hat{F}(t'),\label{eq:0012a1}\\
&&\hat{\mathcal{q}}^{(f)}(t)=\int^{t}_{-\infty}dt'\ G(t-t')\hat{f}(t'),\label{eq:0012a2}
\end{eqnarray}
\end{subequations}
\end{small}and the Green function \begin{small}${G(t)}$\end{small} is given by
\begin{small}
\begin{equation}\label{eq:0013}
G(t)=\frac{1}{2\pi}\int^{\infty}_{-\infty} d\omega \ \alpha(\omega+i0^+)\ e^{-i\omega t},
\end{equation}
\end{small}with the response function
\begin{small}
\begin{equation}\label{eq:0014}
\alpha(z)=\frac{1}{-m_0 z^2-iz\tilde{\mu}(z)+m_0\omega_0^2}.
\end{equation}
\end{small}From Eq. (\ref{eq:0008}) and Eq. (\ref{eq:0012a}), we obtain the correlation
\begin{small}
\begin{equation}\label{eq:0015}
\frac{1}{2}\big\langle \hat{\mathcal{q}}^{(F)}(t) \hat{\mathcal{q}}^{(F)}(t')\!+\!\hat{\mathcal{q}}^{(F)}(t') \hat{\mathcal{q}}^{(F)}(t)\big\rangle=\frac{\hbar}{\pi}\!\int^{\infty}_{0}\!d\omega \ \rm{Im}\{\alpha(\omega+i0^+)\} \coth{\left(\frac{\beta\hbar\omega}{2}\right)}\cos[\omega(t-t')].
\end{equation}
\end{small}\par One can easily find that for negative times the Green function (\ref{eq:0013}) vanishes, and for positive times Eq. (\ref{eq:0013}) is a solution to the following equation,
\begin{small}
\begin{equation}\label{eq:0016}
m_0 \ddot{G}(t)+\int^{t}_{0}\!dt' \mu(t-t')\dot{G}(t')+m_0\omega_0^2 G(t)+\mu(t)G(0)=0,
\end{equation}
\end{small}with the initial conditions \begin{small}${G(0)=0}$\end{small} and \begin{small}${\dot{G}(0)=1/m_0}$\end{small}. Then we obtain the general solutions to the quantum Langevin equation (\ref{eq:0005}) \cite{Ford2001},
\begin{small}
\begin{subequations}\label{eq:0018}
\begin{eqnarray}
&&\hat{\mathcal{q}}(t)=m_0 \dot{G}(t) \hat{\mathcal{q}}(0)+G(t) \hat{\mathcal{p}}(0)+\hat{X}(t)+\hat{Y}(t),\label{eq:0018a} \\
&&\dot{\hat{\mathcal{q}}}(t)=m_0 \ddot{G}(t) \hat{\mathcal{q}}(0)+\dot{G}(t) \hat{\mathcal{p}}(0)+\dot{\hat{X}}(t)+\dot{\hat{Y}}(t),\label{eq:0018b}
\end{eqnarray}
\end{subequations}
\end{small}where \begin{small}${\hat{X}(t)}$\end{small} is the position operator relevant to the fluctuating force \begin{small}${\hat{F}(t)}$\end{small} \cite{Ford2001},
\begin{small}
\begin{equation}\label{eq:0019a}
\hat{X}(t)=\int^{t}_{0}dt'\ G(t-t')\hat{F}(t'),
\end{equation}
\end{small}and \begin{small}${\hat{Y}(t)}$\end{small} is the position operator relevant to the driven force \begin{small}${\hat{f}(t)}$\end{small},
\begin{small}
\begin{equation}\label{eq:0019b}
\hat{Y}(t)=\int^{t}_{0}dt'\ G(t-t')\hat{f}(t').
\end{equation}\end{small}In the following, based on Eqs. (\ref{eq:0018a}) and (\ref{eq:0018b}), we try to obtain the analytical expression of the reduced Wigner function of the system.

\section{Analytical expression of the reduced Wigner function}

In this section, we calculate the reduced Wigner function of the system. By tracing out the degrees of freedom of the heat bath, we get the time-dependent reduced Wigner function of the system \cite{Yu1996},
\begin{small}
\begin{equation}\label{eq:0020}
W(\mathcal{q},\mathcal{p};t)=\int d\mathbf{\boldsymbol{q}}\int d\mathbf{\boldsymbol{p}}\ \it{W}_{\rm{tot}}(\mathcal{q},\mathcal{p};\mathbf{\boldsymbol{q}},\mathbf{\boldsymbol{p}};t)
\end{equation}
\end{small}Here \begin{small}${\it{W}_{\rm{tot}}}$\end{small} is the Wigner function of the composite system, with \begin{small}${\mathbf{\boldsymbol{q}}=(q_1,q_2,...,q_N)}$\end{small} and \begin{small}${\mathbf{\boldsymbol{p}}=(p_1,p_2,...,p_N)}$\end{small} the coordinates and the momenta of the heat bath, respectively. Because the evolution of the composite system satisfies the Liouville-von Neumann equation, we have \cite{Ford2001, Yu1996}
\begin{small}
\begin{equation}\label{eq:0021}
W_{\rm{tot}}(\mathcal{q},\!\mathcal{p};\!\mathbf{\boldsymbol{q}},\!\mathbf{\boldsymbol{p}};t)\!=\!W_{\rm{tot}}(\mathcal{q}(0),\!\mathcal{p}(0);\!\mathbf{\boldsymbol{q}}(0),\!\mathbf{\boldsymbol{p}}(0);0),
\end{equation}
\end{small}where \begin{small}${\mathcal{q}(0)}$\end{small}, \begin{small}${\mathcal{p}(0)}$\end{small}, \begin{small}${\mathbf{\boldsymbol{q}}(0)}$\end{small}, \begin{small}${\mathbf{\boldsymbol{p}}(0)}$\end{small} are the initial values of the coordinates and the momenta, and \begin{small}${\mathcal{q}=\mathcal{q}(t)}$\end{small}, \begin{small}${\mathcal{p}=\mathcal{p}(t)}$\end{small}, \begin{small}${\mathbf{\boldsymbol{q}}=\mathbf{\boldsymbol{q}}(t)}$\end{small}, \begin{small}${\mathbf{\boldsymbol{p}}=\mathbf{\boldsymbol{p}}(t)}$\end{small} are the solutions of the equation of motion (\ref{eq:0003}) and (\ref{eq:0004}). We assume that the initial state of the composite system is in a factorized form, i.e., a direct product of the density operator of the system and that of the heat bath, and the heat bath is in a thermal equilibrium state at the inverse temperature \begin{small}${\beta}$\end{small}. Then we obtain \cite{Ford2001, Yu1996}
\begin{small}
\begin{equation}\label{eq:0022}
W_{\rm{tot}}(\mathcal{q}(0),\mathcal{p}(0);\mathbf{\boldsymbol{q}}(0),\mathbf{\boldsymbol{p}}(0);0)=W(\mathcal{q}(0),\mathcal{p}(0);0)\prod^{N}_{j=1}w_j(q_j(0),p_j(0)),
\end{equation}
\end{small}where \begin{small}${w_j(q_j(0),p_j(0))}$\end{small} is the Wigner function of the \begin{small}${j}$\end{small}th oscillator of the heat bath with the mass \begin{small}${m_j}$\end{small} and the frequency \begin{small}${\omega_j}$\end{small},
\begin{small}
\begin{equation}\label{eq:0023}
w_j(q_j(0),p_j(0))=2\tanh\frac{\beta\hbar\omega_j}{2}\exp\left[-\frac{p_j^2(0)+m^2_j\omega^2_j q_j^2(0)}{m_j\hbar\omega_j\coth(\beta\hbar\omega_j/2)}\right].
\end{equation}
\end{small}Substituting Eqs. (\ref{eq:0021})-(\ref{eq:0023}) into Eq. (\ref{eq:0020}), one obtains
\begin{small}
\begin{equation}\label{eq:0024}
W(\mathcal{q},\mathcal{p};t)=\int d\mathbf{\boldsymbol{q}}(t)\int d\mathbf{\boldsymbol{p}}(t) \it{W}(\mathcal{q}(\rm{0}),\mathcal{p}(\rm{0});0)\it\prod^{N}_{j=\rm{1}}w_j(q_j\rm{(0)},\it{p_j}\rm{(0)}).
\end{equation}
\end{small}Using the general solutions of the quantum Langevin equation (\ref{eq:0018}), we transform the integration variables from the final coordinates of the heat bath \begin{small}${(\mathbf{\boldsymbol{q}}(t), \mathbf{\boldsymbol{p}}(t))}$\end{small} to the initial coordinates of the heat bath \begin{small}${(\mathbf{\boldsymbol{q}}(0), \mathbf{\boldsymbol{p}}(0))}$\end{small}, while holding \begin{small}${\mathcal{q}(0)}$\end{small} and \begin{small}${\mathcal{p}(0)}$\end{small} fixed \cite{Ford2001},
\begin{small}
\begin{equation}\label{eq:0025}
d\mathbf{\boldsymbol{q}}(t)d\mathbf{\boldsymbol{p}}(t)=\frac{1}{m_0^2(\dot{G}^2-G \ddot{G})}d\mathbf{\boldsymbol{q}}(0)d\mathbf{\boldsymbol{p}}(0).
\end{equation}
\end{small}Substituting Eq. (\ref{eq:0025}) into Eq. (\ref{eq:0024}), we obtain\begin{small}
\begin{equation}\label{eq:0026}
W(\mathcal{q},\mathcal{p};t)=\frac{\langle\it{W}(\mathcal{q}(\rm{0}),\mathcal{p}(\rm{0});0)\rangle}{m_0^2(\dot{G}^2-G \ddot{G})},
\end{equation}
\end{small}where the bracket represents the average over the initial equilibrium distribution of the heat bath, and \begin{small}${\mathcal{q}(0)}$\end{small} and \begin{small}${\mathcal{p}(0)}$\end{small} in the integrand can be obtained by inverting Eqs. (\ref{eq:0018a}) and (\ref{eq:0018b}),
\begin{small}
\begin{subequations}\label{eq:0027}
\begin{eqnarray}
&&\mathcal{q}(0)=\frac{m_0\dot{G}(\mathcal{q}-X-Y)-G(\mathcal{p}-m_0\dot{X}-m_0\dot{Y})}{m_0^2(\dot{G}^2-G \ddot{G})}, \label{eq:0027a}\\
&&\mathcal{p}(0)=\frac{-m_0^2\ddot{G}(\mathcal{q}-X-Y)+m_0\dot{G}(\mathcal{p}-m_0\dot{X}-m_0\dot{Y})}{m_0^2(\dot{G}^2-G \ddot{G})}.\label{eq:0027b}
\end{eqnarray}
\end{subequations}
\end{small}We would like to emphasize that in obtaining Eqs. (\ref{eq:0027a}) and (\ref{eq:0027b}), we have performed the Weyl-Wigner transform \cite{Wigner1932, Hillery1984} over Eqs. (\ref{eq:0018a}) and (\ref{eq:0018b}). That is, the operators \begin{small}${\hat{\mathcal{p}}}$\end{small}, \begin{small}${\hat{\mathcal{q}}}$\end{small}, \begin{small}${\hat{\mathcal{p}}(0)}$\end{small}, \begin{small}${\hat{\mathcal{q}}(0)}$\end{small}, \begin{small}${\hat{X}}$\end{small}, \begin{small}${\hat{Y}}$\end{small} in Eqs. (\ref{eq:0018a}) and (\ref{eq:0018b}) have been replaced by variables \begin{small}${\mathcal{p}}$\end{small}, \begin{small}${\mathcal{q}}$\end{small}, \begin{small}${\mathcal{p}(0)}$\end{small}, \begin{small}${\mathcal{q}(0)}$\end{small}, \begin{small}${X}$\end{small}, \begin{small}${Y}$\end{small}. We can calculate the average in Eq. (\ref{eq:0026}) by taking the Fourier transform of the initial reduced Wigner function \cite{Ford2001}
\begin{small}
\begin{equation}\label{eq:0029}
W(\mathcal{q}(0),\mathcal{p}(0);0)\!=\!\frac{1}{(2\pi\hbar)^2}\!\int\!dQ\!\int\!dP\ \tilde{\it{W}}(Q,P;0)\ e^{\frac{i}{\hbar}(\!P \mathcal{q}\!+\!Q \mathcal{p}\!)}.
\end{equation}
\end{small}Inserting this into Eq. (\ref{eq:0026}), after some simplifications, one can obtain \cite{Ford2001}
\begin{small}
\begin{equation}\label{eq:0030}
W(\mathcal{q},\mathcal{p};t)\!=\!\frac{1}{(2\pi\hbar)^2}\!\int^{\infty}_{\!-\infty\!}\!dr\!\!\int^{\infty}_{\!-\infty\!}\!ds
\tilde{\it{W}}\!\!\left(m_0\dot{G}r+Gs, m^2_0\ddot{G}r+\dot{G}s;0\right)\ e^{\frac{i}{\hbar}(r\mathcal{p}+s \mathcal{q}-m_{\rm{0}}\dot{Y}r-Ys)-\frac{1}{2\hbar^2}(m_0^2\langle\dot{X}^2\rangle r^2+m_0\langle X\dot{X}+\dot{X}X\rangle rs+\langle X^2\rangle s^2)}.
\end{equation}
\end{small}Here, we have transformed the integration variables \begin{small}${Q}$\end{small} and \begin{small}${P}$\end{small} into \begin{small}${r}$\end{small} and \begin{small}${s}$\end{small} by
\begin{small}
\begin{equation}\label{eq:0031}
Q=m_0\dot{G}r+Gs,\ \ \ P=m^2_0\ddot{G}r+\dot{G}s,
\end{equation}
\end{small}and we, due to the Gaussian property of \begin{small}${X(t)}$\end{small}, have used \cite{Ford2001}
\begin{small}
\begin{equation}\label{eq:0032}
\langle e^{-\frac{i}{\hbar}(m_0\dot{X}r+Xs)}\rangle\!=\! e^{-\frac{1}{2\hbar^2}(m_0^2\langle\dot{X}^2\rangle r^2+m_0\langle X\dot{X}+\dot{X}X\rangle rs+\langle X^2\rangle s^2)}.
\end{equation}
\end{small}We would like to emphasize that Eq. (\ref{eq:0030}) is the one of the main results of our paper, i.e., the analytical expression of the time evolution of the reduced Wigner function for a quantum Brownian particle in a driven harmonic potential. In this evaluation, the Green function \begin{small}${G(t)}$\end{small} is given by Eq. (\ref{eq:0013}), \begin{small}${X(t)}$\end{small} is given by Eq. (\ref{eq:0019a}) and its correlations are evaluated using Eq. (\ref{eq:0008}), and \begin{small}${Y(t)}$\end{small} is given by Eq. (\ref{eq:0019b}).

Furthermore, one can substitute the inverse of the Fourier transform (\ref{eq:0029}) into Eq. (\ref{eq:0030}) and rewrite the Wigner function in the form of a propagator acting on the initial reduced Wigner function \cite{Ford2001},
\begin{small}
\begin{equation}\label{eq:0033}
W(\mathcal{q},\mathcal{p};t)=\int^{\infty\!}_{-\!\infty\!}\!\frac{d\mathcal{q}'\!(0) d\mathcal{p}'\!(0)}{2\pi\hbar} P(\mathcal{q},\!\mathcal{p};\!\mathcal{q}'\!(0),\!\mathcal{p}'\!(0);\!t)W\!(\mathcal{q}'\!(0),\!\mathcal{p}'\!(0);\!0).
\end{equation}
\end{small}Here the propagator \begin{small}${P(\mathcal{q},\mathcal{p};\mathcal{q}'(0),\mathcal{p}'(0);t)}$\end{small} can be written as \cite{Ford2001}
\begin{small}
\begin{equation}\label{eq:0034}
P(\mathcal{q},\mathcal{p};\mathcal{q}'(0),\mathcal{p}'(0);t)=\frac{\hbar}{\sqrt{|{\bf A}(t)|}}e^{-\frac{1}{2}{\bf R}^{\rm T}(t){\bf A}^{-1}(t){\bf R}(t)},
\end{equation}
\end{small}where \begin{small}${|{\bf A}(t)|}$\end{small} denotes the determinant of \begin{small}${{\bf A}(t)}$\end{small}, and
\begin{small}
\begin{equation}\label{eq:0035}
{\bf A}(t)=\left(
  \begin{array}{cc}
    m_0^2\langle\dot{X}^2\rangle & \frac{m_0}{2}\langle X\dot{X}\!+\!\dot{X}X\rangle\\ \\
    \frac{m_0}{2}\langle X\dot{X}\!+\!\dot{X}X\rangle & \langle X^2\rangle\\
  \end{array}
\right),\ \ \ \ \ \ \ \ \ {\bf R}(t)=\left(
  \begin{array}{c}
    \mathcal{p}-\overline{\mathcal{p}(t)} \\ \\
    \mathcal{q}-\overline{\mathcal{q}(t)} \\
  \end{array}
\right).
\end{equation}
\end{small}Here the overline depicts the average over \begin{small}${X(t)}$\end{small} (\ref{eq:0019a}), and the operators \begin{small}${\overline{\mathcal{q}(t)}}$\end{small} and \begin{small}${\overline{\mathcal{p}(t)}}$\end{small} correspond to the solutions to Eqs. (\ref{eq:0018a}) and (\ref{eq:0018b}) with initial values \begin{small}${\mathcal{q}'(0)}$\end{small} and \begin{small}${\mathcal{p}'(0)}$\end{small},
\begin{small}
\begin{subequations}\label{eq:0037}
\begin{eqnarray}
&&\overline{\mathcal{q}(t)}=m_0\dot{G}(t)\ \mathcal{q}'(0)+G(t)\ \mathcal{p}'(0)+Y(t),\label{eq:0037a} \\
&&\overline{\mathcal{p}(t)}=m_0 \ddot{G}(t)\ \mathcal{q}'(0)+\dot{G}(t)\ \mathcal{p}'(0)+m_0\dot{Y}(t).\label{eq:0037b}
\end{eqnarray}
\end{subequations}
\end{small}

Now we calculate the propagator (\ref{eq:0034}) explicitly. In the case of an Ohmic heat bath, the memory function (\ref{eq:0006}) has the form
\begin{small}
\begin{equation}\label{eq:0038}
\mu(t)=2\gamma_0\delta(t),
\end{equation}
\end{small}where \begin{small}${\gamma_0}$\end{small} is the Newtonian friction constant. Substituting Eq. (\ref{eq:0038}) into Eq. (\ref{eq:0014}), one can obtain
\begin{small}
\begin{equation}\label{eq:0039}
\alpha(z)=\frac{1}{-m_0 z^2-i z \gamma_0+m_0\omega_0^2}.
\end{equation}
\end{small}Substituting Eq. (\ref{eq:0039}) into Eq. (\ref{eq:0013}), after the integration, one can obtain the expression of the Green function. It turns out that in the high damping regime (\begin{small}${\gamma_0/(m_0\omega_0)> 2}$\end{small}),
\begin{small}
\begin{eqnarray}\label{eq:0040}
G(t)=\frac{1}{2m_0\Omega_0}(e^{-\lambda_2 t}-e^{-\lambda_1 t}), \ \ \ \lambda_{1,2}=\frac{\gamma_0}{2m_0}\pm\Omega_0,\\ \nonumber
\end{eqnarray}
\end{small}and in the low damping regime (\begin{small}${\gamma_0/(m_0\omega_0)< 2}$\end{small}),
\begin{small}
\begin{eqnarray}\label{eq:0041}
G(t)=\frac{1}{2i m_0\Omega_0}(e^{-\lambda_2 t}-e^{-\lambda_1 t}), \ \ \ \lambda_{1,2}=\frac{\gamma_0}{2m_0}\pm i\Omega_0. \\ \nonumber
\end{eqnarray}
\end{small}where
\begin{small}
\begin{equation}\label{eq:0041a}
\Omega_0=\sqrt{\left|\frac{\gamma_0^2}{4m_0^2}-\omega_0^2\right|}.
\end{equation}
\end{small}It is straightforward to prove that in both the high and the low damping regimes,
\begin{small}
\begin{equation}\label{eq:0042}
G(-t')-G(t-t')=G(-t')-m_0 G(-t')\left[\dot{G}(t)+\frac{\xi}{2}G(t)\right]-m_0 G(t)\left[\dot{G}(-t')+\frac{\gamma_0}{2m_0}G(-t')\right].
\end{equation}
\end{small}From Eq. (\ref{eq:0012a1}) and Eq. (\ref{eq:0019a}), after performing the Weyl-Wigner transform, we have
\begin{small}
\begin{equation}\label{eq:0043}
X\!(t)\!=\!\mathcal{q}^{(\!F\!)}(t)\!-\!\mathcal{q}^{(\!F\!)}(0)\!+\!\int^{0}_{\!-\infty\!}\!dt' \left[G(-t')\!-\!G(t-t')\right]\!F(t').
\end{equation}
\end{small}Substituting Eq. (\ref{eq:0042}) into Eq. (\ref{eq:0043}), one obtains
\begin{small}
\begin{equation}\label{eq:0044}
X\!(t)\!=\!\mathcal{q}^{(\!F\!)}\!(t)-m_0\!\left[\!\dot{G}(t)\!+\!\frac{\gamma_0}{m_0} G(t)\!\right]\!\mathcal{q}^{(\!F\!)}\!(0)-m_0 G(t)\dot{\mathcal{q}}^{(\!F\!)}\!(t).
\end{equation}
\end{small}From the Weyl-Wigner transform of the correlations of \begin{small}${\mathcal{q}^{(\!F\!)}\!(t)}$\end{small}, i.e., Eq. (\ref{eq:0015}), one can obtain the elements of the matrix of \begin{small}${{\bf A}(t)}$\end{small} \cite{Ford2001} as follows,
\begin{small}
\begin{subequations}
\begin{eqnarray}
&&\langle X^2\rangle\!=\!\frac{1}{2}\left\{1+m_0^2\left[\dot{G}(t)+\frac{\gamma_0}{m_0} G(t)\right]^2\right\}s(0)-\frac{1}{2}m_0^2 G^2(t) \ddot{s}(0) -m_0\!\!\left[\dot{G}(t)+\frac{\gamma_0}{m_0} G(t)\right]\!\!s(t)+m_0 G(t) \dot{s}(t),
\label{eq:0045a}\\
&&\langle \dot{X}^2\rangle\!=\!-\frac{1}{2}\left[1+m_0^2\dot{G}^2(t)\right]\ddot{s}(0)+\frac{1}{2}m_0^2 \!\!\left[\ddot{G}(t)+\frac{\gamma_0}{m_0} \dot{G}(t)\right]^2\!\! s(0)-m_0\!\!\left[\ddot{G}(t)+\frac{\gamma_0}{m_0} \dot{G}(t)\right]\!\!\dot{s}(t)+m_0 \dot{G}(t) \ddot{s}(t),\label{eq:0045b} \\
&&\langle X\dot{X}\!\!+\!\!\dot{X}X\rangle\!=\!m_0^2\!\!\left[\!\dot{G}(t)\!\!+\!\!\frac{\gamma_0}{m_0}\! G(t)\!\right]\!\!\!\left[\!\ddot{G}(t)\!\!+\!\!\frac{\gamma_0}{m_0}\! \dot{G}(t)\!\right]\!\!s(0)\!-\!m_0^2 G(t) \dot{G}(t)\ddot{s}(0)\!-\!m_0\!\!\left[\!\ddot{G}(t)\!\!+\!\!\frac{\gamma_0}{m_0}\! \dot{G}(t)\!\right]\!\!s(t)\!-\!\gamma_0 G(t)\dot{s}(t)\!+\!m_0 G(t) \ddot{s}(t),\ \ \ \ \ \ \ \ \ \ \ \ \ \label{eq:0045c}
\end{eqnarray}
\end{subequations}
\end{small}where
\begin{small}
\begin{equation}\label{eq:0046}
s(t)\!=\!\frac{2\gamma_0}{\pi}\!\int^{\infty}_{0}\!d\omega \frac{\hbar\omega}{m_0^2(\omega_0^2-\omega^2)^2\!+\!\gamma_0^2\omega_0^2}\coth{\frac{\beta\hbar\omega}{2}}\cos{\omega t}.
\end{equation}
\end{small}Note that Eqs. (\ref{eq:0045a})-(\ref{eq:0045c}) are valid in both the high and the low damping regimes.

Before proceeding to the next step, let us make a self-consistency check about our results of Eqs. (\ref{eq:0045a})-(\ref{eq:0045c}). It is expected that in the high temperature limit, Eqs. (\ref{eq:0045a})-(\ref{eq:0045c}) will reproduce their classical counterparts (see Sec.10.2.1 in Ref. \cite{Risken1996}). In the high temperature limit, \begin{small}${\coth{(\beta\hbar\omega/2)}\to 2/(\beta\hbar\omega)}$\end{small}, then Eq. (\ref{eq:0046}) becomes
\begin{small}
\begin{equation}\label{eq:0046a}
s_{\rm cl}(t)\!=\!\frac{4\gamma_0}{\beta\pi}\!\int^{\infty}_{0}\!d\omega \frac{\cos{\omega t}}{m_0^2(\omega_0^2-\omega^2)^2\!+\!\gamma_0^2\omega_0^2}.
\end{equation}
\end{small}After taking the integration, one can obtain
\begin{small}
\begin{equation}\label{eq:0047}
s_{\rm cl}(t)\!=\!\left\{
\begin{aligned}
&\frac{1}{\beta m_0 \Omega_0 \omega_0^2}(\!-\lambda_2 e^{-\lambda_1 t}\!+\!\lambda_1 e^{-\lambda_2 t}),\ \ \ \ \ \frac{\gamma_0}{m_0\omega_0}\!\!>\!\!2, \\
&\frac{1}{i\beta m_0 \Omega_0 \omega_0^2}(\!-\lambda_2 e^{-\lambda_1 t}\!+\!\lambda_1 e^{-\lambda_2 t}),\ \ \ \ \frac{\gamma_0}{m_0\omega_0}\!\!<\!\!2.
\end{aligned}
\right.\\ \
\end{equation}
\end{small}Substituting Eqs. (\ref{eq:0040}), (\ref{eq:0041}), and (\ref{eq:0047}) into Eqs. (\ref{eq:0045a})-(\ref{eq:0045c}), after some simplification, one can obtain the classical limit of Eq. (\ref{eq:0035}),
\begin{small}
\begin{equation}\label{eq:0047a}
{\bf A}_{\rm cl}(t)=\left(
  \begin{array}{cc}
    m_0^2\langle\dot{X}^2\rangle_{\rm cl} & \frac{m_0}{2}\langle X\dot{X}\!+\!\dot{X}X\rangle_{\rm cl} \\ \\
    \frac{m_0}{2}\langle X\dot{X}\!+\!\dot{X}X\rangle_{\rm cl} & \langle X^2\rangle_{\rm cl} \\
  \end{array}
\right),
\end{equation}
\end{small}where
\begin{small}
\begin{subequations}\label{eq:0048}
\begin{eqnarray}
&&\langle X^2\rangle_{\rm cl}=\frac{\gamma_0 v^2_{th}}{m_0(\lambda_1-\lambda_2)^2}\!\!\left[\frac{\lambda_1+\lambda_2}{\lambda_1\lambda_2}+\frac{4}{\lambda_1+\lambda_2}(e^{-(\lambda_1+\lambda_2)t}\!-\!1)\!-\!\frac{1}{\lambda_1}e^{-2\lambda_1 t}\!-\!\frac{1}{\lambda_2}e^{-2\lambda_2 t}\right], \label{eq:0048a} \\
&&m_0^2\langle \dot{X}^2\rangle_{\rm cl}=\frac{m_0^2\gamma_0 v^2_{th}}{m_0(\lambda_1-\lambda_2)^2}\!\!\left[\lambda_1+\lambda_2+\frac{4\lambda_1\lambda_2}{\lambda_1+\lambda_2}(e^{-(\lambda_1+\lambda_2)t}\!-\!1)\!-\!\lambda_1 e^{-2\lambda_1 t}\!-\!\lambda_2 e^{-2\lambda_2 t}\right], \label{eq:0048b}\\
&&\frac{m_0}{2}\langle X\dot{X}\!+\!\dot{X}X\rangle_{\rm cl}=\frac{m_0 \gamma_0 v^2_{th}}{(\lambda_1\!-\!\lambda_2)^2}(e^{-\lambda_1 t}\!-\!e^{-\lambda_2 t})^2,\label{eq:0048c}
\end{eqnarray}
\end{subequations}
\end{small}and \begin{small}${v^2_{th}=1/(\beta m_0)}$\end{small}. One can see that Eqs. (\ref{eq:0048a})-(\ref{eq:0048c}) are exactly the same as Eq. (10.63) in Ref. \cite{Risken1996}, i.e., our results (\ref{eq:0045a})-(\ref{eq:0045c}) reproduce the results of the classical Brownian particle in the high temperature limit (please note that ${m_0}$ has been set to one and \begin{small}${f(t)=0}$\end{small} in Eq. (10.63) in Ref. \cite{Risken1996}). Substituting Eqs. (\ref{eq:0035}), (\ref{eq:0037a})-(\ref{eq:0037b}), (\ref{eq:0046a}) and (\ref{eq:0048a})-(\ref{eq:0048c}) into Eq. (\ref{eq:0034}), one obtains the classical propagator in the phase space, which is the same as Eq. (10.55) in Ref. \cite{Risken1996}:
\begin{small}
\begin{equation}\label{eq:0049}
P_{\!\rm cl}(\mathcal{q},\!\mathcal{p};\!\mathcal{q}'\!(0),\!\mathcal{p}'\!(0);\!t)\!=\!\frac{\hbar}{\!\sqrt{|\!{\bf A}_{\rm cl}\!(t)\!|}}\!\exp\!\left[\!-\frac{1}{2}[{\bf A}_{\rm cl}^{\!-\!1\!}(t)]_{\!pp}[\mathcal{p}\!-\!\overline{\mathcal{p}\!(t)}]^2\!-\![{\bf A}_{\rm cl}^{\!-\!1\!}(t)]_{\!pq}[\mathcal{p}\!-\!\overline{\mathcal{p}\!(t)}][\mathcal{q}\!-\!\overline{\mathcal{q}\!(t)}]\!-\!\frac{1}{2}[{\bf A}_{\rm cl}^{\!-\!1\!}(t)]_{\!qq}[\mathcal{q}\!-\!\overline{\mathcal{q}\!(t)}]^2\! \right].
\end{equation}
\end{small}Here, the elements of the inverse matrix of \begin{small}${{\bf A}_{\rm cl}(t)}$\end{small} are given by
\begin{small}
\begin{subequations}
\begin{eqnarray}
&&[{\bf A}_{\rm cl}^{\!-\!1\!}(t)]_{pp}=\langle X^2\rangle_{\rm cl}/|\!{\bf A}_{\rm cl}\!(t)\!|, \\
&&[{\bf A}_{\rm cl}^{\!-\!1\!}(t)]_{pq}=[{\bf A}_{\rm cl}^{\!-\!1\!}(t)]_{qp}\!=\!-m_0\langle X\dot{X}\!+\!\dot{X}X\rangle_{\rm cl}/(2|\!{\bf A}_{\rm cl}\!(t)\!|),\\
&&[{\bf A}_{\rm cl}^{\!-\!1\!}(t)]_{qq}=m_0^2\langle \dot{X}^2\rangle_{\rm cl}/|\!{\bf A}_{\rm cl}\!(t)\!|,
\end{eqnarray}
\end{subequations}
\end{small}and
\begin{small}
\begin{equation}\label{eq:0050d}
|{\bf A}_{\rm cl}\!(t)|\!=\!m_0^2\left[\langle X^2\rangle_{\rm cl} \! \langle\dot{X}^2\rangle_{\rm cl}\!-\!\langle X\dot{X}\!+\!\dot{X}X\rangle_{\rm cl}^2/4\right].
\end{equation}
\end{small}

We now take the thermal equilibrium initial state as an example to calculate the time evolution of the Wigner function of the system (Eqs. (\ref{eq:0033}) and (\ref{eq:0034})). Please note that the initial state can also be any state other than the thermal equilibrium state. But for simplicity, we use the thermal equilibrium state as an example to demonstrate the effectiveness of our method. We assume that the system is initially prepared in a thermal equilibrium state at the inverse temperature \begin{small}${\beta'}$\end{small}, which is different from the temperature of the heat bath \begin{small}${\beta}$\end{small}, and the Wigner function of the initial density matrix can be written as
\begin{small}
\begin{equation}\label{eq:0051}
W({\mathcal{p}'}\!(0),{ \mathcal{q}'}\!(0);0)\!=\!2\tanh\!{\frac{\beta'\!\hbar\omega_0}{2}}\exp\!\!\left[\!-\frac{1}{2}\!\left(\mathcal{p}'\!(0)\ \ \mathcal{q}'\!(0)\right)\!{\bf B}\!\left(\!
  \begin{array}{c}
    \mathcal{p}'\!(0) \\ \\
    \mathcal{q}'\!(0)
  \end{array}
\!\right)\!\right]\!,\ \ \
{\bf B}\!=\!\!\left(\!
  \begin{array}{cc}
    \!2\left(\!m_0\hbar\omega_0\coth{\frac{\beta'\!\hbar\omega_0}{2}}\!\right)^{\!-1\!}\! & 0 \\
    0 & \!2\left(\!\frac{\hbar}{m_0\omega_0}\coth{\frac{\beta'\!\hbar\omega_0}{2}}\!\right)^{\!-1\!}\!\\
  \end{array}
\!\right)\!.
\end{equation}
\end{small}Substituting Eqs. (\ref{eq:0034}) and (\ref{eq:0051}) into Eq. (\ref{eq:0033}), one can obtain
\begin{small}
\begin{eqnarray}\label{eq:0053}
W(\mathcal{p},\mathcal{q};t)&=&\frac{2\tanh{\frac{\beta'\hbar\omega}{2}}}{\sqrt{|{\bf A}(t)||{\bf B}+{\bf \Lambda}^{\rm T}(t){\bf A}^{-1}(t){\bf \Lambda}(t)|}}\exp\left[\frac{1}{2}\left(\mathcal{p}-m_0\dot{Y}(t)\ \ \ \mathcal{q}-Y(t) \right)\right.\nonumber \\
&&\left. \cdot\left\{{\bf A}^{-1}(t){\bf \Lambda}(t)[{\bf B}+{\bf \Lambda}^{\rm T}(t){\bf A}^{-1}(t){\bf \Lambda}(t)]^{-1}{\bf \Lambda}^{\rm T}(t){\bf A}^{-1}(t)-{\bf A}^{-1}(t)\right\}\cdot\left(
  \begin{array}{c}
    \mathcal{p}-m_0\dot{Y}(t) \\
    \mathcal{q}-Y(t)
  \end{array}
\right)\right],
\end{eqnarray}\end{small}where
\begin{small}
\begin{equation}\label{eq:0054}
{\bf \Lambda}(t)=\left(
  \begin{array}{cc}
    m_0\dot{G}(t) & m_0^2\ddot{G}(t) \\ \\
    G(t) & m_0\dot{G}(t) \\
  \end{array}
\right).
\end{equation}
\end{small}From this result, one can easily find that the variances of the Wigner function is independent of the external force \begin{small}${f(t)}$\end{small}.

Furthermore, we would like to show how the equilibrium solution arises in the long time limit, i.e., the relaxation process from \begin{small}${T'=1/\beta'}$\end{small} to \begin{small}${T=1/\beta}$\end{small}:

First we recall that, so long as the angular frequency of the system \begin{small}${\omega_0}$\end{small} is nonzero, the Green function will vanish as \begin{small}${t\to\infty}$\end{small} \cite{Ford2001}, thus \begin{small}${{\bf \Lambda}(t)={\bf 0}}$\end{small} when \begin{small}${t\to\infty}$\end{small}.

Next, from Eqs. (\ref{eq:0045a})-(\ref{eq:0045c}), we have
\begin{small}
\begin{equation}\label{eq:0055}
\langle X^2\rangle=\frac{1}{2}s(0),\ \ \ \langle \dot{X}^2\rangle=-\frac{1}{2}\ddot{s}(0),\ \ \ \langle X\dot{X}\!+\!\dot{X}X\rangle=0.
\end{equation}
\end{small}

For simplicity, we consider the weak coupling limit. In the weak coupling limit,
\begin{small}
\begin{equation}\label{eq:0056}
\lim_{\gamma_0\to0}\frac{2\gamma_0}{\pi} \frac{\hbar\omega}{m_0^2(\omega_0^2-\omega^2)^2\!+\!\gamma_0^2\omega_0^2}=\frac{\hbar\omega}{m_0\omega_0^2}\delta(\omega-\omega_0).
\end{equation}
\end{small}Then we have
\begin{small}
\begin{equation}\label{eq:0057}
{\bf A}(t)=\left(
  \begin{array}{cc}
    \frac{m_0\hbar\omega_0}{2}\coth{\frac{\beta\hbar\omega_0}{2}} & 0 \\ \\
    0 & \frac{\hbar}{2m_0\omega_0}\coth{\frac{\beta\hbar\omega_0}{2}} \\
  \end{array}
\right).
\end{equation}
\end{small}Substituting Eqs. (\ref{eq:0054}), (\ref{eq:0055}), and (\ref{eq:0057}) into Eq. (\ref{eq:0053}), one can obtain the asymptotic expression of the Wigner function in the long time limit
\begin{small}
\begin{equation}\label{eq:0058}
W(\mathcal{q},\mathcal{p};t)=2\tanh{\frac{\beta\hbar\omega_0}{2}}\exp\!\!\left[-\frac{[\mathcal{p}-m_0\dot{Y}(t)]^2}{m_0\hbar\omega_0\coth{\frac{\beta\hbar\omega_0}{2}}}-\frac{[\mathcal{q}-Y(t)]^2}{\frac{\hbar}{m_0\omega_0}\coth{\frac{\beta\hbar\omega_0}{2}}}\right].
\end{equation}
\end{small}This is the familiar form of the Wigner function of a dragged harmonic oscillator, which is independent of the initial temperature \begin{small}${\beta'}$\end{small}. Please note that when the driving force vanishes, i.e., \begin{small}${\hat{f}(t)=0}$\end{small}, our result (\ref{eq:0058}) reproduces the result in Ref. \cite{Ford2001}.

\section{Quantum corrections to the entropy of the system}
Based on the above results, we now calculate the quantum corrections to the entropy of a dragged harmonic oscillator which is undergoing quantum Brownian motion. First, we assume that the initial state of the system has a well-defined classical counterpart \cite{Qiu2020}, i.e., when we expand the initial Wigner function in powers of \begin{small}${\hbar}$\end{small}, there are no terms in negative powers of \begin{small}${\hbar}$\end{small},
\begin{small}
\begin{equation}\label{eq:0059}
W(\mathcal{q}'\!(0),\mathcal{p}'\!(0);0)=W_{\rm cl}+(i\hbar) W^{(1)}+(i\hbar)^2 W^{(2)}+o(\hbar^2),
\end{equation}
\end{small}where \begin{small}${W_{\rm cl}(\mathcal{q}'\!(0),\mathcal{p}'\!(0);0)}$\end{small} is the corresponding classical probability distribution in the phase space. Next, we expand the propagator (\ref{eq:0034}) in powers of \begin{small}${\hbar}$\end{small}. Because
\begin{small}
\begin{equation}\label{eq:0060}
\hbar\coth{\frac{\beta\hbar\omega}{2}}=\frac{2}{\beta\omega}-(i\hbar)^2\frac{\beta\omega}{6}+o(\hbar^2),
\end{equation}
\end{small}we have
\begin{small}
\begin{equation}\label{eq:0061}
s(t)=s_{\rm cl}(t)+(i\hbar)^2 s^{(2)}(t)+o(\hbar^2),
\end{equation}
\end{small}where \begin{small}${s^{(2)}(t)}$\end{small} is given by
\begin{small}
\begin{equation}\label{eq:0062}
s^{(2)}(t)\!=\!-\frac{\beta\gamma_0}{3\pi}\!\int^{\infty}_{0}\!d\omega \frac{\omega^2\cos{\omega t}}{m_0^2(\omega_0^2-\omega^2)^2\!+\!\gamma_0^2\omega_0^2}.
\end{equation}
\end{small}After taking the integration, one can obtain
\begin{small}
\begin{equation}\label{eq:0062a}
s^{(2)}(t)=\left\{
\begin{aligned}
&\frac{\beta}{12 m_0 \Omega_0}(\lambda_a e^{-\lambda_1 t}\!-\!\lambda_2 e^{-\lambda_2 t}),\ \ \ \frac{\gamma_0}{m_0\omega_0}\!>\!2, \\
&\frac{-i\beta}{12 m_0 \Omega_0}(\lambda_1 e^{-\lambda_1 t}\!-\!\lambda_2 e^{-\lambda_2 t}),\ \ \ \frac{\gamma_0}{m_0\omega_0}\!<\!2.
\end{aligned}
\right.\\ \
\end{equation}
\end{small}From Eqs. (\ref{eq:0045a})-(\ref{eq:0045c}), we know that
\begin{small}
\begin{eqnarray}
\langle X^2\rangle\!\!&=&\!\!\langle X^2\rangle_{\rm cl}+(i\hbar)^2\langle X^2\rangle^{(2)}+o(\hbar^2),\label{eq:0063a}\\
\langle \dot{X}^2\rangle\!\!&=&\!\!\langle \dot{X}^2\rangle_{\rm cl}+(i\hbar)^2\langle \dot{X}^2\rangle^{(2)}+o(\hbar^2),\label{eq:0063b}\\
\langle X\dot{X}\!+\!\dot{X}X\rangle\!\!&=&\!\!\langle X\dot{X}\!+\!\dot{X}X\rangle_{\rm cl}\!+\!(i\hbar)^2\langle X\dot{X}\!+\!\dot{X}X\rangle^{\!(2)\!}\!+\!o(\hbar^2),\label{eq:0063c}
\end{eqnarray}
\end{small}where the expressions of \begin{small}${\langle X^2\rangle^{(2)}}$\end{small}, \begin{small}${\langle \dot{X}^2\rangle^{(2)}}$\end{small} and \begin{small}${\langle X\dot{X}\!+\!\dot{X}X\rangle^{\!(2)\!}}$\end{small} can be obtained by substituting Eq. (\ref{eq:0062}) into Eqs. (\ref{eq:0045a})-(\ref{eq:0045c}). Then the matrix \begin{small}${{\bf A}(t)}$\end{small} can be expanded in powers of \begin{small}${\hbar}$\end{small} as follows,
\begin{small}
\begin{equation}\label{eq:0064}
{\bf A}(t)={\bf A}_{\rm cl}(t)+(i\hbar)^2 {\bf A}^{(2)}(t)+o(\hbar^2),
\end{equation}
\end{small}where \begin{small}${{\bf A}^{(2)}(t)}$\end{small} is given by
\begin{small}
\begin{equation}\label{eq:0065}
{\bf A}^{(2)}(t)=\left(
  \begin{array}{cc}
    m_0^2\langle\dot{X}^2\rangle^{\!(2)\!} & \frac{m_0}{2}\langle X\dot{X}\!+\!\dot{X}X\rangle^{\!(2)\!}\\ \\
    \frac{m_0}{2}\langle X\dot{X}\!+\!\dot{X}X\rangle^{\!(2)\!} & \langle X^2\rangle^{\!(2)\!} \\
  \end{array}
\right).
\end{equation}
\end{small}Then we can similarly expand \begin{small}${|{\bf A}(t)|}$\end{small} and \begin{small}${{\bf A}^{-1}(t)}$\end{small} in powers of \begin{small}${\hbar}$\end{small},
\begin{small}
\begin{subequations}
\begin{eqnarray}
|{\bf A}(t)|&=&|{\bf A}_{\rm cl}(t)|+(i\hbar)^2|{\bf A}_{\rm cl}(t)|\ {\rm Tr}[{\bf A}_{\rm cl}^{-1}(t){\bf A}^{(2)}(t)]+o(\hbar^2),\label{eq:0066a}\\
{\bf A}^{-1}(t)&=&{\bf A}_{\rm cl}^{-1}(t)-(i\hbar)^2{\bf A}_{\rm cl}^{-1}(t){\bf A}^{(2)}(t){\bf A}_{\rm cl}^{-1}(t)+o(\hbar^2).\label{eq:0066b}
\end{eqnarray}
\end{subequations}
\end{small}

Substituting Eqs. (\ref{eq:0066a}) and (\ref{eq:0066b}) into Eq. (\ref{eq:0034}), we obtain the expression of the propagator in powers of \begin{small}${\hbar}$\end{small} as follows,
\begin{small}
\begin{equation}\label{eq:0067}
P(\mathcal{q},\mathcal{p};\mathcal{q}'\!(0),\mathcal{p}'\!(0);t)=P_{\rm cl}(\mathcal{q},\mathcal{p};\mathcal{q}'\!(0),\mathcal{p}'\!(0);t)+(i\hbar)^2 P^{(2)}(\mathcal{q},\mathcal{p};\mathcal{q}'\!(0),\mathcal{p}'\!(0);t)+o(\hbar^2),
\end{equation}
\end{small}where \begin{small}${P_{\rm cl}(\mathcal{q},\mathcal{p};\mathcal{q}'\!(0),\mathcal{p}'\!(0);t)}$\end{small} is given by Eq. (\ref{eq:0049}), and
\begin{small}
\begin{equation}\label{eq:0068}
P^{(2)}(\mathcal{q},\mathcal{p};\mathcal{q}'\!(0),\mathcal{p}'\!(0);t)\!=\!\frac{1}{2}P_{\rm cl}(\mathcal{q},\mathcal{p};\mathcal{q}'\!(0),\mathcal{p}'\!(0);t)\left({\bf R}^{\rm T}(t){\bf A}_{\rm cl}^{-1}(t){\bf A}^{(2)}(t){\bf A}_{\rm cl}^{-1}(t){\bf R}(t)-{\rm Tr}[{\bf A}_{\rm cl\!}^{\!-\!1\!}(t){\bf A}^{\!(2)\!}(t)]\right).
\end{equation}
\end{small}Substituting Eqs. (\ref{eq:0059}) and (\ref{eq:0067}) into Eqs. (\ref{eq:0033}), we obtain the time evolution of the Wigner function of the system in powers of \begin{small}${\hbar}$\end{small},
\begin{small}
\begin{equation}\label{eq:0069}
W(\mathcal{q},\mathcal{p};t)=W_{\rm cl}(\mathcal{q},\mathcal{p};t)+(i\hbar)W^{(1)}(\mathcal{q},\mathcal{p};t)+(i\hbar)^2 W^{(2)}(\mathcal{q},\mathcal{p};t)+o(\hbar^2),
\end{equation}
\end{small}where
\begin{small}
\begin{subequations}
\begin{eqnarray}
&&\!\!\!W_{\rm cl}(\mathcal{q},\mathcal{p};t)\!=\!\int^{\infty}_{-\infty\!}\!\frac{d\mathcal{q}'\!(0) d\mathcal{p}'\!(0)}{2\pi\hbar} P_{\rm cl}(\mathcal{q},\mathcal{p};\mathcal{q}'\!(0),\mathcal{p}'\!(0);t)W_{\rm cl}(\mathcal{q}'\!(0),\mathcal{p}'\!(0);0),\label{eq:0070a} \\
&&\!\!\!W^{(1)}(\mathcal{q},\mathcal{p};t)\!=\!\int^{\infty}_{-\infty\!}\!\frac{d\mathcal{q}'\!(0) d\mathcal{p}'\!(0)}{2\pi\hbar} P_{\rm cl}(\mathcal{q},\mathcal{p};\mathcal{q}'\!(0),\mathcal{p}'\!(0);t)W^{(1)}(\mathcal{q}'\!(0),\mathcal{p}'\!(0);0),\label{eq:0070b}\\
&&\!\!\!W^{\!(2)\!}(\mathcal{q},\mathcal{p};t)\!=\!\int^{\infty\!}_{\!-\!\infty\!}\!\frac{d\mathcal{q}'\!(0) d\mathcal{p}'\!(0)}{2\pi\hbar}\! \left[\!P_{\rm cl}(\mathcal{q},\!\mathcal{p};\!\mathcal{q}'\!(0),\!\mathcal{p}'\!(0);t)W^{\!(2)\!}(\mathcal{q}'\!(0),\mathcal{p}'\!(0);0)\!+\!P^{(2)\!}(\mathcal{q},\!\mathcal{p};\!\mathcal{q}'\!(0),\!\mathcal{p}'\!(0);\!t)W_{\rm cl}(\mathcal{q}'\!(0),\mathcal{p}'\!(0);0)\!\right]\!.\ \ \ \ \ \ \ \ \ \ \label{eq:0070c}
\end{eqnarray}
\end{subequations}
\end{small}One can see that \begin{small}${W_{\rm cl}(\mathcal{q},\mathcal{p};t)}$\end{small} is the corresponding classical probability distribution at time \begin{small}${t}$\end{small} in the phase space, while \begin{small}${W^{(1)}(\mathcal{q},\mathcal{p};t)}$\end{small} and \begin{small}${W^{(2)}(\mathcal{q},\mathcal{p};t)}$\end{small} are the first- and the second-order quantum corrections at time \begin{small}${t}$\end{small} to the classical probability distribution, respectively. It is worth mentioning that this result is a dynamical extension to results in Ref. \cite{Wigner1932}.

Finally, using the methods developed in Ref. \cite{Qiu2020}, we obtain the quantum corrections to the classical Gibbs entropy of a dragged harmonic oscillator which is undergoing quantum Brownian motion,
\begin{small}
\begin{equation}\label{eq:0071}
S_q(t)=S_{\rm cl}(t)+(i\hbar)S^{(1)}(t)+(i\hbar)^2 S^{(2)}(t)+o(\hbar^2),
\end{equation}
\end{small}
where
\begin{small}
\begin{subequations}
\begin{eqnarray}
S_{\rm cl}(t)\!&=&\!-\int\frac{d\mathcal{q}d\mathcal{p}}{2\pi\hbar}\ W_{\rm cl} \ln W_{\rm cl}.\label{eq:0072a}\\
S^{(1)}(t)\!&=&\!-\int\frac{d\mathcal{q}d\mathcal{p}}{2\pi\hbar}\ \left[W^{(1)} \ln W_{\rm cl}+W^{(1)}\right],\label{eq:0072b}\\
S^{(2)}(t)\!&=&\!-\!\int\!\frac{d\mathcal{q}d\mathcal{p}}{2\pi\hbar}\ \left[W^{(2)} \ln W_{\rm cl}\!+\!W^{(2)}
\!+\!\frac{(W^{(1)})^2}{2W_{\rm cl}}-\frac{W_{\rm cl}(\stackrel{\leftarrow}{\partial_{\mathcal{p}}}\stackrel{\rightarrow}{\partial_{\mathcal{q}}}
-\stackrel{\leftarrow}{\partial_{\mathcal{q}}}\stackrel{\rightarrow}{\partial_{\mathcal{p}}})^2W_{\rm cl}}{16W_{\rm cl}}+\frac{G(W_{\rm cl})}{12W_{\rm cl}^2}\right].\label{eq:0072c}
\end{eqnarray}
\end{subequations}
\end{small}Here,
\begin{small}
\begin{equation}\label{eq:0073}
G(W_{\rm cl})=(\partial^2_{\mathcal{p}} W_{\rm cl})(\partial_{\mathcal{q}} W_{\rm cl})^2+(\partial^2_{\mathcal{q}} W_{\rm cl})(\partial_{\mathcal{p}} W_{\rm cl})^2-2(\partial_{\mathcal{q}} W_{\rm cl})(\partial_{\mathcal{p}} W_{\rm cl})(\partial_{\mathcal{q}\mathcal{p}} W_{\rm cl}).
\end{equation}
\end{small}One can find that \begin{small}${S_{\rm cl}(t)}$\end{small} is exactly the corresponding classical Gibbs entropy of the system, while \begin{small}${S^{(1)}(t)}$\end{small} and \begin{small}${S^{(2)}(t)}$\end{small} are the first- and the second-order quantum corrections to the entropy, respectively.

As a demonstration, we take the thermal equilibrium initial state as an example to show our results (\ref{eq:0071})-(\ref{eq:0073}). We assume that the system is prepared initially in the thermal equilibrium state at the inverse temperature \begin{small}${\beta'}$\end{small}. The initial Wigner function is given by Eq. (\ref{eq:0051}), which can be expanded in the form of Eq. (\ref{eq:0059}) and
\begin{small}
\begin{subequations}\label{eq:0074}
\begin{eqnarray}
&&W_{\rm cl}(\mathcal{q}'(0),\mathcal{p}'(0);0)=\beta'\hbar\omega_0e^{-\beta'\epsilon(\mathcal{q}'(0),\mathcal{p}'(0))},\label{eq:0074a}\\
&&W^{(1)}(\mathcal{q}'(0),\mathcal{p}'(0);0)=0,\label{eq:0074b}\\
&&W^{\!(2)\!}(\mathcal{q}'\!(0),\!\mathcal{p}'\!(0);\!0)\!=\!W_{\rm\! cl}\frac{{\beta'}^2\!\omega_0^2}{12}\![1\!-\!\beta'\!\epsilon(\mathcal{q}'\!(0),\!\mathcal{p}'\!(0))],\label{eq:0074c}
\end{eqnarray}
\end{subequations}
\end{small}where
\begin{small}
\begin{equation}\label{eq:0075}
\epsilon(\mathcal{q}'\!(0),\mathcal{p}'\!(0))=\frac{{\mathcal{p}'}^2\!(0)}{2m_0}+\frac{1}{2}m_0\omega_0^2{\mathcal{q}'}^2\!(0)
\end{equation}
\end{small}is the Hamiltonian of a single harmonic oscillator. Substituting Eqs. (\ref{eq:0059}), (\ref{eq:0067}), (\ref{eq:0068}), and (\ref{eq:0074a})-(\ref{eq:0074c}) into Eq. (\ref{eq:0069}), one can obtain the expansion of the reduced Wigner function at an arbitrary time \begin{small}${t}$\end{small}. Finally, we obtain the the quantum corrections to the entropy by substituting Eqs. (\ref{eq:0070a})-(\ref{eq:0070c}) into Eqs. (\ref{eq:0072a})-(\ref{eq:0072c}). One can find that for the thermodynamic equilibrium initial state, all terms odd in \begin{small}${\hbar}$\end{small} are exactly zero due to \begin{small}${W^{(1)}(\mathcal{q}'\!(0),\mathcal{p}'\!(0);0)=0}$\end{small}. The evolution of the classical Wigner function \begin{small}${W_{\rm cl}(\mathcal{q},\mathcal{p};t)}$\end{small} is given by Eq. (\ref{eq:0070a}), and \begin{small}${S_{\rm cl}(t)}$\end{small} reproduces the classical Gibbs entropy for a dragged Brownian harmonic oscillator. And the lowest order quantum correction to the entropy of the system is given by Eq. (\ref{eq:0072c}).

\section{Quantum corrections to the total entropy production rate and the heat dissipation rate}
Based on the results (\ref{eq:0072a})-(\ref{eq:0072c}) in Sec. V, we can get more information and physical insights about the entropy production in a nonequilibrium quantum stochastic process. In Refs. \cite{Ge2009, Ge2010, Esposito2010b, Esposito2010c}, the expressions of the classical total entropy production rate \begin{small}${e^{({\rm cl})}_{\rm p}}$\end{small} and the classical heat dissipation rate \begin{small}${h^{({\rm cl})}_{\rm d}}$\end{small} are defined in a nonequilibrium stochastic process. Then the second law is reformulated in the nonequilibrium form, and the total entropy production rate can be further split into two nonnegative parts,
\begin{small}
\begin{equation}\label{eq:0076}
e^{({\rm cl})}_{\rm p}=\beta Q^{({\rm cl})}_{\rm hk}+(-\beta f^{({\rm cl})}_{\rm d}).
\end{equation}
\end{small}Here, \begin{small}${Q^{({\rm cl})}_{\rm hk}}$\end{small} is the classical housekeeping heat, which represents the irreversible work done by the surrounding to the system that is kept away from reaching equilibrium, and \begin{small}${f^{({\rm cl})}_{\rm d}}$\end{small} is the classical free energy dissipation rate associated with spontaneous relaxation \cite{Ge2010}. Nevertheless, a quantum version of the total entropy production rate and the heat dissipation rate has largely been unexplored so far, except for the open quantum systems described by the Lindblad master equation \cite{Spohn1978a, Spohn1978b}. Based on the results in Sec. V, we now derive the expression of the quantum corrections to the total entropy production rate and the heat dissipation rate.

We consider an Ohmic heat bath with the memory function (\ref{eq:0038}). Under the Markovian approximation, the time evolution of the reduced Wigner function of the system satisfies the Caldeira-Leggett master equation \cite{Caldeira1983, Hu1992, Yu1996, Ford2001}, which has the same form as the Kramers equation \cite{Risken1996}
\begin{small}
\begin{equation}\label{eq:0077}
\partial_t W(\mathcal{q},\mathcal{p};t)=-\frac{\mathcal{p}}{m_0}\partial_\mathcal{q}\!W(\mathcal{q},\mathcal{p};t)\!+\!m_0\omega_0^2\left(\!\mathcal{q}(t)\!-\!\frac{f(t)}{m_0\omega_0^2}\!\right)\partial_\mathcal{p}\!W(\mathcal{q},\mathcal{p};t)+\frac{\gamma_0}{m_0}\partial_\mathcal{p}\!\left[\mathcal{p}W(\mathcal{q},\mathcal{p};t)\right]+\frac{\gamma_0}{\beta}\partial^2_{\mathcal{p}}W(\mathcal{q},\mathcal{p};t).
\end{equation}
\end{small}In the overdamped limit (\begin{small}${\gamma_0\gg 1}$\end{small}), the solution to Eq. (\ref{eq:0077}) can be written as \cite{Risken1996}
\begin{small}
\begin{equation}\label{eq:0078} W(\mathcal{q},\mathcal{p};t)=w(\mathcal{q};t)\mathcal{W}^{(eq)}(\mathcal{p}),
\end{equation}
\end{small}Here, \begin{small}${\mathcal{W}^{(eq)}(\mathcal{p})}$\end{small} is the  Wigner function in the momentum space which is always in the thermal equilibrium state,
\begin{small}
\begin{equation}\label{eq:0079}
\mathcal{W}^{(eq)}(\mathcal{p})\!=\!\sqrt{\frac{\beta}{2\pi m_0}}\exp{\!\left(\!-\beta\frac{\mathcal{p}^2}{2 m_0}\!\right)},
\end{equation}
\end{small}and \begin{small}${w(\mathcal{q};t)}$\end{small} is the Wigner function in the coordinate space which satisfies the Smoluchowski equation \cite{Ge2009, Esposito2010c},
\begin{small}
\begin{equation}\label{eq:0080}
\partial_t w(\mathcal{q};t)=-\partial_{\mathcal{q}}j(\mathcal{q};t),
\end{equation}
\end{small}where\begin{small}
\begin{equation}\label{eq:0081}
j(\mathcal{q};t)=b(\mathcal{q};t)w(\mathcal{q};t)-\frac{1}{\beta\gamma_0}\partial_{\mathcal{q}}w(\mathcal{q};t)
\end{equation}
\end{small}is the probability flux and\begin{small}
\begin{equation}\label{eq:0082}
b(\mathcal{q};t)=\frac{1}{\gamma_0}\left[-m_0\omega_0^2\mathcal{q}+f(t)\right]
\end{equation}
\end{small}is the drift coefficient.

In order to derive the quantum corrections to the total entropy production rate and the heat dissipation rate, we expand \begin{small}${e_{\rm p}}$\end{small}, \begin{small}${h_{\rm d}}$\end{small}, and \begin{small}${j(\mathcal{q};t)}$\end{small} in powers of \begin{small}${\hbar}$\end{small} as
\begin{small}
\begin{eqnarray}
e_{\rm p}&=&e^{({\rm cl})}_{\rm p}+(i\hbar)e^{(1)}_{\rm p}+(i\hbar)^2 e^{(2)}_{\rm p}+o(\hbar^2),\label{eq:0083a}\\
h_{\rm d}&=&h^{({\rm cl})}_{\rm d}+(i\hbar)h^{(1)}_{\rm d}+(i\hbar)^2 h^{(2)}_{\rm d}+o(\hbar^2),\label{eq:0083b}\\
j&=&j_{\rm cl}+(i\hbar)j^{(1)}+(i\hbar)^2 j^{(2)}+o(\hbar^2),\label{eq:0083c}
\end{eqnarray}
\end{small}and Eqs. (\ref{eq:0080}-\ref{eq:0081}) become
\begin{small}
\begin{subequations}
\begin{equation}\label{eq:0084}
\left\{
\begin{aligned}
&\partial_t w_{\rm cl}(\mathcal{q};t)=-\partial_{\mathcal{q}}j_{\rm cl}(\mathcal{q};t), \\
&j_{\rm cl}(\mathcal{q};t)=b(\mathcal{q};t)w_{\rm cl}(\mathcal{q};t)-\frac{1}{\beta\gamma_0}\partial_{\mathcal{q}}w_{\rm cl}(\mathcal{q};t),
\end{aligned}
\right.\ \ \ \ \ \
\end{equation}
\begin{equation}\label{eq:0085}
\left\{
\begin{aligned}
&\partial_t w^{(1)}(\mathcal{q};t)=-\partial_{\mathcal{q}}j^{(1)}(\mathcal{q};t), \\
&j^{(1)}(\mathcal{q};t)=b(\mathcal{q};t)w^{(1)}(\mathcal{q};t)-\frac{1}{\beta\gamma_0}\partial_{\mathcal{q}}w^{(1)}(\mathcal{q};t),
\end{aligned}
\right.
\end{equation}
\begin{equation}\label{eq:0086}
\left\{
\begin{aligned}
&\partial_t w^{(2)}(\mathcal{q};t)=-\partial_{\mathcal{q}}j^{(2)}(\mathcal{q};t), \\
&j^{(2)}(\mathcal{q};t)=b(\mathcal{q};t)w^{(2)}(\mathcal{q};t)-\frac{1}{\beta\gamma_0}\partial_{\mathcal{q}}w^{(2)}(\mathcal{q};t).
\end{aligned}
\right.
\end{equation}
\end{subequations}
\end{small}According to Refs. \cite{Qian2001, Qian2001b, Seifert2005}, the quantum heat dissipation rate can be written as
\begin{small}
\begin{equation}\label{eq:0086a}
h_{\rm d}=\beta\gamma_0 \int\frac{d\mathcal{q}d\mathcal{p}}{2\pi\hbar}b(\mathcal{q};t)J(\mathcal{q},\mathcal{p};t),
\end{equation}
\end{small}where \begin{small}${J(\mathcal{q},\mathcal{p};t)}$ ${=}$ ${j(\mathcal{q};t)\mathcal{W}^{(eq)}}(\mathcal{p})$\end{small}. By substituting Eq. (\ref{eq:0083c}) into Eq. (\ref{eq:0086a}), one can obtain
\begin{small}
\begin{subequations}
\begin{eqnarray}
&&h^{({\rm cl})}_{\rm d}=\beta\gamma_0 \int\frac{d\mathcal{q}d\mathcal{p}}{2\pi\hbar}b(\mathcal{q};t)J_{\rm cl}(\mathcal{q},\mathcal{p};t),\label{eq:0089}\\
&&h^{(1)}_{\rm d}=\beta\gamma_0 \int\frac{d\mathcal{q}d\mathcal{p}}{2\pi\hbar}b(\mathcal{q};t)J^{(1)}(\mathcal{q},\mathcal{p};t),\label{eq:0090}\\
&&h^{(2)}_{\rm d}=\beta\gamma_0 \int\frac{d\mathcal{q}d\mathcal{p}}{2\pi\hbar}b(\mathcal{q};t)J^{(2)}(\mathcal{q},\mathcal{p};t),\label{eq:0091}
\end{eqnarray}
\end{subequations}
\end{small}where \begin{small}${J_{\rm cl}(\mathcal{q},\mathcal{p};t)}$ ${=}$ ${j_{\rm cl}(\mathcal{q};t)\mathcal{W}^{(eq)}(\mathcal{p})}$\end{small}, \begin{small}${J^{(1)}(\mathcal{q},\mathcal{p};t)}$ ${=}$ ${j^{(1)}(\mathcal{q};t)}$ ${\mathcal{W}^{(eq)}(\mathcal{p})}$\end{small} and \begin{small}${J^{(2)}(\mathcal{q},\mathcal{p};t)}$ ${=}$ ${j^{(2)}(\mathcal{q};t)}$ ${\mathcal{W}^{(eq)}(\mathcal{p})}$\end{small}. The quantum total entropy production rate is given by
\begin{small}
\begin{equation}\label{eq:0087}
e_{\rm p}\equiv\frac{d S(t)}{dt}-h_{\rm d}.
\end{equation}
\end{small}By taking the derivative of \begin{small}${S_{\rm cl}(t)}$\end{small}, \begin{small}${S^{(1)}(t)}$\end{small} and \begin{small}${S^{(2)}(t)}$\end{small} in Eqs. (\ref{eq:0072a})-(\ref{eq:0072c}) and using the dynamical equations (\ref{eq:0084})-(\ref{eq:0086}), one can obtain
\begin{small}
\begin{subequations}
\begin{eqnarray}
e^{({\rm cl})}_{\rm p}&=&\beta\gamma_0 \int\frac{d\mathcal{q}d\mathcal{p}}{2\pi\hbar}\frac{J^2_{\rm cl}(\mathcal{q},\mathcal{p};t)}{W_{\rm cl}(\mathcal{q},\mathcal{p};t)},\label{eq:0088}\\
e^{(1)}_{\rm p}&=&2\beta\gamma_0\!\! \int\!\frac{d\mathcal{q}d\mathcal{p}}{2\pi\hbar}\frac{J^{(1)}(\mathcal{q},\mathcal{p};t)J_{\rm cl}(\mathcal{q},\mathcal{p};t)}{W_{\rm cl}(\mathcal{q},\mathcal{p};t)}-\beta\gamma_0\!\! \int\!\frac{d\mathcal{q}d\mathcal{p}}{2\pi\hbar}\frac{J^2_{\rm cl}(\mathcal{q},\mathcal{p};t)}{W^2_{\rm cl}(\mathcal{q},\mathcal{p};t)}W^{(1)}(\mathcal{q},\mathcal{p};t),\label{eq:0092}\\
e^{(2)}_{\rm p}&=&2\beta\gamma_0\!\! \int\!\frac{d\mathcal{q}d\mathcal{p}}{2\pi\hbar}\frac{J^{(2)}(\mathcal{q},\mathcal{p};t)J_{\rm cl}(\mathcal{q},\mathcal{p};t)}{W_{\rm cl}(\mathcal{q},\mathcal{p};t)}\!-\!\beta\gamma_0\!\! \int\!\frac{d\mathcal{q}d\mathcal{p}}{2\pi\hbar}\frac{J^2_{\rm cl}(\mathcal{q},\mathcal{p};t)}{W^2_{\rm cl}(\mathcal{q},\mathcal{p};t)}W^{(2)}(\mathcal{q},\mathcal{p};t)\nonumber\\
&&+\beta\gamma_0\!\int\!\frac{d\mathcal{q}d\mathcal{p}}{2\pi\hbar}\left[\frac{J^{(1)}(\mathcal{q},\mathcal{p};t)W_{\rm cl}(\mathcal{q},\mathcal{p};t)-W^{(1)}(\mathcal{q},\mathcal{p};t)J_{\rm cl}(\mathcal{q},\mathcal{p};t)}{W_{\rm cl}(\mathcal{q},\mathcal{p};t)}\right]^2\frac{1}{W_{\rm cl}(\mathcal{q},\mathcal{p};t)}\nonumber\\
&&+\int\!\frac{d\mathcal{q}d\mathcal{p}}{2\pi\hbar}\frac{\partial}{\partial t}\!\!\left[\frac{W_{\rm cl}(\mathcal{q},\mathcal{p};t)(\stackrel{\leftarrow}{\partial_{\mathcal{p}}}\stackrel{\rightarrow}{\partial_{\mathcal{q}}}
-\stackrel{\leftarrow}{\partial_{\mathcal{q}}}\stackrel{\rightarrow}{\partial_{\mathcal{p}}})^2W_{\rm cl}(\mathcal{q},\mathcal{p};t)}{16W_{\rm cl}(\mathcal{q},\mathcal{p};t)}\right]-\!
\int\!\frac{d\mathcal{q}d\mathcal{p}}{2\pi\hbar}\frac{\partial}{\partial t}\!\!\left[\frac{G(W_{\rm cl}(\mathcal{q},\mathcal{p};t))}{12W_{\rm cl}^2(\mathcal{q},\mathcal{p};t)}\right].\label{eq:0093}
\end{eqnarray}
\end{subequations}
\end{small}Here, \begin{small}${h^{({\rm cl})}_{\rm d}}$\end{small}
is the heat dissipation rate and \begin{small}${e^{({\rm cl})}_{\rm p}}$\end{small} is the nonnegative classical entropy production rate. One can see that Eq. (\ref{eq:0089}) and Eq. (\ref{eq:0088}) are exactly the same as the results in Ref. \cite{Ge2009, Esposito2010c}, thus our results reproduce their classical counterparts in the classical limit (\begin{small}${\hbar\to 0}$\end{small}).

By rewriting Eq. (\ref{eq:0000}) in the phase space formulation, one can obtain the exact expression of the von Neumann entropy \cite{Zachos2007},
\begin{small}
\begin{equation}\label{eq:0094a}
S(t)=-\int\frac{d\mathcal{q}d\mathcal{p}}{2\pi\hbar}\ W(\mathcal{q},\mathcal{p};t) \ln_\star W(\mathcal{q},\mathcal{p};t),
\end{equation}
\end{small}where
\begin{small}
\begin{equation}\label{eq:0094b}
\ln_\star W(\mathcal{q},\mathcal{p};t)\equiv-\sum^{\infty}_{n=1}\frac{[1-W(\mathcal{q},\mathcal{p};t)]^n_\star}{n},
\end{equation}
\end{small}and the Moyal product \cite{Hillery1984}
\begin{small}
\begin{equation}\label{eq:0094c}
\star\equiv\exp{\left[-\frac{i\hbar}{2}(\stackrel{\leftarrow}{\partial_{\mathcal{p}}}\stackrel{\rightarrow}{\partial_{\mathcal{q}}}
-\stackrel{\leftarrow}{\partial_{\mathcal{q}}}\stackrel{\rightarrow}{\partial_{\mathcal{p}}})\right]}.
\end{equation}
\end{small}If we expand Eq. (\ref{eq:0094c}) in powers of \begin{small}${\hbar}$\end{small} and take the lowest-order approximation of the Moyal product, the von Neumann entropy is approximated as \cite{Santos2017}
\begin{small}
\begin{equation}\label{eq:0094}
\mathcal{S}(t)=-\int\frac{d\mathcal{q}d\mathcal{p}}{2\pi\hbar}\ W(\mathcal{q},\mathcal{p};t) \ln W(\mathcal{q},\mathcal{p};t),
\end{equation}
\end{small}and then the approximate expression of the quantum entropy production rate \begin{small}${\mathcal{e}_{\rm p}}$\end{small} and the quantum heat dissipation rate \begin{small}${\mathcal{h}_{\rm d}}$\end{small} can be written as
\begin{small}
\begin{equation}\label{eq:0095}
\mathcal{e}_{\rm p}=\beta\gamma_0 \int\frac{d\mathcal{q}d\mathcal{p}}{2\pi\hbar}\frac{J^2(\mathcal{q},\mathcal{p};t)}{W(\mathcal{q},\mathcal{p};t)},\ \ \ \ \ \ \ \
\end{equation}
\begin{equation}\label{eq:0096}
\mathcal{h}_{\rm d}=\beta\gamma_0 \int\frac{d\mathcal{q}d\mathcal{p}}{2\pi\hbar}b(\mathcal{q};t)J(\mathcal{q},\mathcal{p};t),
\end{equation}
\end{small}where \begin{small}${J(\mathcal{q},\mathcal{p};t)}$ ${=}$ ${j(\mathcal{q};t)}$ ${\mathcal{W}^{(eq)}(\mathcal{p})}$\end{small}. Moreover, one can split \begin{small}${\mathcal{e}_{\rm p}}$\end{small} into two parts \cite{Ge2009, Esposito2010c},
\begin{small}
\begin{equation}\label{eq:0097}
\mathcal{e}_{\rm p}=\beta\mathcal{Q}_{\rm hk}+(-\beta\mathcal{f}_{\rm d}),
\end{equation}
\end{small}where \begin{small}${\mathcal{Q}_{\rm hk}}$\end{small} is the quantum housekeeping heat which is given by
\begin{small}
\begin{equation}\label{eq:0097a}
\mathcal{Q}_{\rm hk}\!=\!\gamma_0\!\int\!\frac{d\mathcal{q}d\mathcal{p}}{2\pi\hbar}\!\left[b(\mathcal{q};t)\!-\!\frac{1}{\beta\gamma_0}\frac{\partial\!\log{W_{\!\rm st}(\mathcal{q},\mathcal{p};t)}}{\partial\mathcal{q}}\right]\!W(\mathcal{q},\mathcal{p};t),
\end{equation}
\end{small}and \begin{small}${\mathcal{f}_{\rm d}}$\end{small} is the quantum free energy dissipation rate which is given by
\begin{small}
\begin{equation}\label{eq:0097b}
\mathcal{f}_{\rm d}\!=\!\frac{1}{\beta^2\gamma_0}\!\int\!\!\frac{d\mathcal{q}d\mathcal{p}}{2\pi\hbar}\!\!\left[\frac{\partial}{\partial\mathcal{q}}\!\log\frac{W_{\rm st}(\mathcal{q},\mathcal{p};t)}{W(\mathcal{q},\mathcal{p};t)}\right]^2\!\!W(\mathcal{q},\mathcal{p};t).
\end{equation}
\end{small}Here \begin{small}${W_{\rm st}(\mathcal{q},\mathcal{p};t)}$\end{small} is the reduced Wigner function of the steady state of the system, i.e., the long time limit (\begin{small}${t\to\infty}$\end{small}) of \begin{small}${W(\mathcal{q},\mathcal{p};t)}$\end{small}. Note that Eqs. (\ref{eq:0097})-(\ref{eq:0097b}) are not the exact results. They are valid only under the approximation (\ref{eq:0094}). Moreover, by expanding \begin{small}${\mathcal{e}_{\rm p}}$\end{small} in powers of \begin{small}${\hbar}$\end{small}, we obtain
\begin{small}
\begin{equation}\label{eq:0098}
\mathcal{e}_{\rm p}=\mathcal{e}^{({\rm cl})}_{\rm p}+(i\hbar)\mathcal{e}^{(1)}_{\rm p}+(i\hbar)^2 \mathcal{e}^{(2)}_{\rm p}+o(\hbar^2),
\end{equation}
\end{small}where
\begin{small}
\begin{subequations}
\begin{eqnarray}
&&\!\!\!\!\!\!\!\!\mathcal{e}^{({\rm cl})}_{\rm p}\!=\!\beta\gamma_0 \int\frac{d\mathcal{q}d\mathcal{p}}{2\pi\hbar}\frac{J^2_{\rm cl}(\mathcal{q},\mathcal{p};t)}{W_{\rm cl}(\mathcal{q},\mathcal{p};t)},\label{eq:0099}\\
&&\!\!\!\!\!\!\!\!\mathcal{e}^{(1)}_{\rm p}\!=\!2\beta\gamma_0\!\! \int\!\frac{d\mathcal{q}d\mathcal{p}}{2\pi\hbar}\frac{J^{(1)}(\mathcal{q},\mathcal{p};t)J_{\rm cl}(\mathcal{q},\mathcal{p};t)}{W_{\rm cl}(\mathcal{q},\mathcal{p};t)}\!-\!\beta\gamma_0\!\! \int\!\frac{d\mathcal{q}d\mathcal{p}}{2\pi\hbar}\frac{J^2_{\rm cl}(\mathcal{q},\mathcal{p};t)}{W^2_{\rm cl}(\mathcal{q},\mathcal{p};t)}W^{(1)}(\mathcal{q},\mathcal{p};t),\label{eq:0100}\\
&&\!\!\!\!\!\!\!\!\mathcal{e}^{(2)}_{\rm p}\!=\!2\beta\gamma_0\!\! \int\!\frac{d\mathcal{q}d\mathcal{p}}{2\pi\hbar}\frac{J^{(2)}(\mathcal{q},\mathcal{p};t)J_{\rm cl}(\mathcal{q},\mathcal{p};t)}{W_{\rm cl}(\mathcal{q},\mathcal{p};t)}\!-\!\beta\gamma_0\!\! \int\!\frac{d\mathcal{q}d\mathcal{p}}{2\pi\hbar}\frac{J^2_{\rm cl}(\mathcal{q},\mathcal{p};t)}{W^2_{\rm cl}(\mathcal{q},\mathcal{p};t)}W^{(2)}(\mathcal{q},\mathcal{p};t)\!+\!\beta\gamma_0\!\!\int\!\!\frac{d\mathcal{q}d\mathcal{p}}{2\pi\hbar}\!\!\left[\!\frac{J^{(1)}W_{\rm cl}\!-\!W^{(1)}J_{\rm cl}}{W_{\rm cl}}\!\right]^{\!2}\!\!\frac{1}{W_{\rm cl}}.\nonumber\\ \label{eq:0101}
\end{eqnarray}
\end{subequations}
\end{small}One can see that the lowest and the first order quantum correction of \begin{small}${\mathcal{e}_{\rm p}}$\end{small} (Eq. (\ref{eq:0099}) and Eq. (\ref{eq:0100})) are the same as Eq. (\ref{eq:0088}) and Eq. (\ref{eq:0092}), respectively. The second order quantum correction of \begin{small}${\mathcal{e}_{\rm p}}$\end{small} (\ref{eq:0101}) reproduces the first three terms while misses the last two terms of Eq. (\ref{eq:0093}). The last two terms just arise from the Moyal product when we rewrite Eq. (\ref{eq:0000}) in the phase space formulation \cite{Qiu2020}.

Furthermore, as a demonstration, we take the thermal equilibrium initial state and the linear external force as an example to show our results (\ref{eq:0088})-(\ref{eq:0093}). We assume that the system is prepared initially in the thermal equilibrium state at the inverse temperature \begin{small}${\beta'}$\end{small}. The initial Wigner function is given by Eq. (\ref{eq:0051}). The linear external force can be written as \begin{small}
\begin{equation}\label{eq:0102}
f(t)=m_0\omega_0^2\cdot vt,
\end{equation}
\end{small}where \begin{small}${v}$\end{small} is the velocity of the driving. Under the Markovian approximation and in the overdamped limit, the solution to the Wigner function of the system can be obtained from Eq. (\ref{eq:0033}),
\begin{small}
\begin{equation}\label{eq:0103}
w(\mathcal{q};t)\!=\!\frac{2\pi\hbar}{\sqrt{\frac{2\pi}{\beta m_0\omega_0^2}\!\left[1\!+\!\left(\!\frac{\beta\hbar\omega_0}{2}\!\coth{\frac{\beta'\hbar\omega_0}{2}}\!-\!1\!\right)\exp{\!\left(\!-2\frac{m_0\omega_0^2}{\gamma_0}t\right)}\right]}}\exp\!{\left\{\!-\frac{\left[\mathcal{q}-vt+\frac{v\gamma_0}{m_0\omega_0^2}\left(1-\exp{\left(\!-\frac{m_0\omega_0^2}{\gamma_0}t\right)}\right)\right]^2}{\frac{2}{\beta m_0\omega_0^2}\!\!\left[1\!+\!\left(\!\frac{\beta\hbar\omega_0}{2}\!\coth{\frac{\beta'\hbar\omega_0}{2}}\!-\!1\!\right)\exp{\!\left(\!-2\frac{m_0\omega_0^2}{\gamma_0}t\right)}\right]}\right\}}.
\end{equation}
\end{small}By expanding Eq. (\ref{eq:0079}) and Eq. (\ref{eq:0103}) in powers of \begin{small}${\hbar}$\end{small}, one can obtain
\begin{small}
\begin{subequations}
\begin{eqnarray}
&&\!\!\!\!\!\!\!\!\!\!W_{\rm cl}(\mathcal{q},\mathcal{p};t)\!=\!\frac{2\pi\hbar\ \mathcal{W}^{(eq)}(\mathcal{p})}{\sqrt{\frac{2\pi}{\beta m_0\omega_0^2}\!\left[1\!+\!\left(\!\frac{\beta}{\beta'}\!-\!1\!\right)\exp{\!\left(\!-2\frac{m_0\omega_0^2}{\gamma_0}t\right)}\right]}}\exp\!{\left\{\!-\frac{\left[\mathcal{q}-vt+\frac{v\gamma_0}{m_0\omega_0^2}\left(1-\exp{\left(\!-\frac{m_0\omega_0^2}{\gamma_0}t\right)}\right)\right]^2}{\frac{2}{\beta m_0\omega_0^2}\!\!\left[1\!+\!\left(\!\frac{\beta}{\beta'}\!-\!1\!\right)\exp{\!\left(\!-2\frac{m_0\omega_0^2}{\gamma_0}t\right)}\right]}\right\}},\label{eq:0103a}\\
&&\!\!\!\!\!\!\!\!\!\!W^{(1)}(\mathcal{q},\mathcal{p};t)=0,\ \ \ \ \ \ \ \ \ \ \ \ \ \label{eq:0103b}\\
&&\!\!\!\!\!\!\!\!\!\!W^{(2)}(\mathcal{q},\mathcal{p};t)=0.\label{eq:0103c}
\end{eqnarray}
\end{subequations}
\end{small}Note that all the quantum correction terms of \begin{small}${W(\mathcal{q},\mathcal{p};t)}$\end{small} are exactly zero because all the correction terms of \begin{small}${\mathcal{W}^{(eq)}(\mathcal{p})}$\end{small} are zero. By substituting Eqs. (\ref{eq:0103a})-(\ref{eq:0103c}) into Eqs. (\ref{eq:0088})-(\ref{eq:0093}), we obtain the quantum corrections to the total entropy production rate and the heat dissipation rate. One can see that the even order correction terms of \begin{small}${e_{\rm p}}$\end{small} are nonzero due to the last two terms in Eq. (\ref{eq:0093}). Therefore, the Moyal product has played an important role in the calculation of the quantum corrections to the entropy production rate.

From the above analysis, one can see that under the Markovian approximation, the quantum corrections to the total entropy production rate \begin{small}${e_{\rm p}(t)}$\end{small} and the heat dissipation rate \begin{small}${h_{\rm d}(t)}$\end{small} arise from two different physical origins. One is the difference between the definition of the von Neumann entropy and Gibbs entropy (the Moyal product), the other is the difference between the initial Wigner function and the initial classical probability distribution.




\section{Dicussion and Summary}
Before concluding this paper, we would like to give the following remarks.

(I) In calculating the von Neumann entropy, we trace out the degrees of freedom of the heat bath, and ignore completely the entanglement between the system and the heat bath. It can be seen that when neglecting the entanglement, the von Neumann entropy of the system reproduces its classical counterpart in the classical limit (\ref{eq:0072a}). However, it is unclear to us if it is proper to neglect the entanglement in the study of quantum information related problems, e.g., the Landauer's principle \cite{Horhammer2008, Horhammer2005}. How the entanglement between the system and the heat bath will influence the Landauer's principle in an open quantum system is still an open question.

(II) Exactly solvable models can bring important insights. The dynamics of the quantum Brownian motion model under a time-dependent Hamiltonian is of great importance in the study of nonequilibrium quantum thermodynamics, e.g., finite time quantum heat engines, quantum Landauer's principle, and quantum fluctuation theorems. But it is usually extremely difficult to solve exactly due to the huge number of degrees of freedom of the heat bath. Luckily, for this specific model, we obtain the analytical results of the time evolution of the Wigner function and the von Neumann entropy. The exact solutions of the quantum corrections to the entropy will be helpful for analyzing the interplay between quantum mechanics and thermodynamics at extremely low temperature.

(III) We also notice that in Refs. \cite{Agarwal1971a, Agarwal1971b, Agarwal1973}, the author presented a method to calculate the von Neumann entropy of quantum states whose Wigner function is in a Gaussian form. So this method can also be applied to calculate the entropy of Eq. (\ref{eq:0053}). However, their method is not applicable when the Wigner function is non-Gaussian. Nevertheless, the method for calculating the von Neumann entropy presented in Eqs. (\ref{eq:0071})-(\ref{eq:0073}) is valid for whatever states as long as they have well-defined classical counterparts.

(IV) Our results about the quantum correction to the total entropy production rate \begin{small}${e_{\rm p}(t)}$\end{small} and the heat dissipation rate \begin{small}${h_{\rm d}(t)}$\end{small} are valid under the Markovian approximation. In Ref. \cite{R2012}, the authors extend the definition of the classical nonadiabatic entropy production given in Ref. \cite{Esposito2010a} to an arbitrary non-Markovian systems. We plan to extend our results (\ref{eq:0088})-(\ref{eq:0093}) to arbitrary non-Markovian stochastic process in the future.

In summary, in this paper, we study the time evolution of the von Neumann entropy of a quantum Brownian particle under a driving force, as well as the total entropy production. By solving the quantum Langevin equation, we obtain the analytical expression of the Wigner function at an arbitrary time \begin{small}${t}$\end{small}. As an example, we obtain the evolution of the Wigner function explicitly when the system is initially prepared in a thermal equilibrium state, and it reproduces the classical probability distribution in the high-temperature and the weak-coupling limit. Based on the above results and the results of the \begin{small}${\hbar}$\end{small} expansion of the von Neumann entropy in the phase space, we prove that the zeroth-order term reproduces the Gibbs entropy, and we obtain the explicit expression of the time evolution of the quantum corrections to the Gibbs entropy. Moreover, under the Markovian approximation, we obtain the expression of the quantum corrections to the total entropy production rate \begin{small}${e_{\rm p}(t)}$\end{small} and the heat dissipation rate \begin{small}${h_{\rm d}(t)}$\end{small}.

In the classical stochastic thermodynamics, fluctuating work (heat) is defined along individual stochastic trajectory in the phase space \cite{Sekimoto2010}. Nevertheless, it is elusive to define a trajectory-dependent work (heat) in open quantum systems, because there is no well-defined trajectory in the Hilbert space due to the Heisenberg uncertainty principle. We plan to extend our current investigation to these problems and we believe that further studies along this line will advance our understanding about the relationship between the quantum and the classical work and heat and may bring important insights to some fundamental problems in quantum thermodynamics.

\section{Acknowledgment}
H. T. Quan acknowledges support from the National Science Foundation of China under grants 11775001, 11534002, and 11825001.

\end{document}